\pdfoutput=1
\documentclass[conference,compsoc]{IEEEtran}
\IEEEoverridecommandlockouts
\usepackage[T1]{fontenc}
\usepackage{cite}
\usepackage{booktabs,array,tabularx,multirow,colortbl}
\usepackage{amsmath,amssymb,amsthm}
\usepackage{graphicx,xcolor}
\newcommand{\pThreeTable}{%
  \arrayrulecolor{black}%
  \fontsize{7.8bp}{8.9bp}\selectfont
  \renewcommand{\arraystretch}{1}%
  \setlength{\heavyrulewidth}{.6bp}%
  \setlength{\lightrulewidth}{.3bp}%
  \setlength{\aboverulesep}{1.1bp}%
  \setlength{\belowrulesep}{1.1bp}%
  \setlength{\defaultaddspace}{1.5bp}%
}
\newcommand{\pThreeHead}[1]{{\bfseries #1}}
\newcommand{\pThreeSmallHead}[1]{{\fontsize{7.5bp}{8.2bp}\selectfont\bfseries #1}}
\newcommand{\pThreeMark}{\raisebox{.15ex}{\scalebox{.62}{\ding{108}}}}

\usepackage{listings}
\usepackage{tikz}
\usetikzlibrary{arrows.meta,positioning,fit,calc}
\usepackage[font=footnotesize,labelsep=colon]{caption}
\usepackage[hidelinks]{hyperref}
\hypersetup{pdftitle={When Answering Is Administering: Permission Replies That Rewrite Coding-Agent Policy}}
\usepackage{url}
\expandafter\def\expandafter\UrlBreaks\expandafter{\UrlBreaks\do\-}
\usepackage{balance}
\usepackage{pifont}
\usepackage{enumitem}
\newcommand{\circnum}[1]{\ding{\numexpr181+#1\relax}}
\renewcommand{\paragraph}[1]{\par\noindent\textbf{#1.}\hspace{0.5em}\ignorespaces}

\definecolor{lost}{HTML}{B2182B}
\definecolor{kept}{HTML}{2166AC}
\definecolor{neutral}{HTML}{59636F}
\definecolor{losttint}{HTML}{FDDBC7}
\definecolor{kepttint}{HTML}{D1E5F0}
\definecolor{neutraltint}{HTML}{F3F4F6}
\definecolor{ink}{HTML}{20242C}
\definecolor{p3band}{HTML}{FEEDE5}
\definecolor{p3rule}{HTML}{AAB3BC}
\definecolor{p3fail}{HTML}{FDDBC7}
\definecolor{gridc}{HTML}{E3E7EC}
\definecolor{lostbar}{HTML}{F4A582}
\definecolor{keptbar}{HTML}{92C5DE}
\definecolor{neutralbar}{HTML}{C9CED4}

\newtheorem{theorem}{Theorem}
\newtheorem{corollary}{Corollary}
\newtheorem{lemma}{Lemma}
\newtheorem{proposition}{Proposition}
\theoremstyle{definition}
\newtheorem{definition}{Definition}

\newcommand{\M}{\mathit{M}}
\newcommand{\code}[1]{\textup{\texttt{#1}}}

\definecolor{lost}{HTML}{B2182B}
\definecolor{kept}{HTML}{2166AC}
\definecolor{neutral}{HTML}{59636F}
\definecolor{losttint}{HTML}{FDDBC7}
\definecolor{kepttint}{HTML}{D1E5F0}
\definecolor{neutraltint}{HTML}{F3F4F6}
\definecolor{ink}{HTML}{20242C}
\definecolor{p3band}{HTML}{FEEDE5}
\definecolor{p3rule}{HTML}{AAB3BC}
\definecolor{p3fail}{HTML}{FDDBC7}
\definecolor{gridc}{HTML}{E3E7EC}
\definecolor{lostbar}{HTML}{F4A582}
\definecolor{keptbar}{HTML}{92C5DE}
\definecolor{neutralbar}{HTML}{C9CED4}
\newcommand{\Pbody}{\fontsize{8bp}{8.7bp}\selectfont}

\newcommand{\Pparty}[1]{\textit{e}\kern-.2bp\raisebox{-1.7bp}{\fontsize{7bp}{7bp}\selectfont #1}}
\newcommand{\Pcopy}{\textit{X}\kern-.3bp\raisebox{1.8bp}{\fontsize{7bp}{7bp}\selectfont$\prime$}}
\tikzset{
 ptext/.style={inner sep=0bp,outer sep=0bp,align=center,text=black,font=\Pbody},
 parrow/.style={draw=neutral,line width=.65bp,-{Latex[length=2bp,width=1.6bp]}},
 rarrow/.style={draw=lost,line width=.75bp,-{Latex[length=2bp,width=1.6bp]}},
 barrow/.style={draw=kept,line width=.65bp,dash pattern=on 2.2bp off 1.7bp,-{Latex[length=2bp,width=1.6bp]}},
 pbox/.style={draw=neutral,line width=.55bp,rounded corners=2bp,fill=white}}

\renewcommand{\tablename}{Table}
\begin{document}

\title{When Answering Is Administering:\\Permission Replies That Rewrite Coding-Agent Policy}
\author{\IEEEauthorblockN{Zeming Fei\IEEEauthorrefmark{1},
Hongming Fei\IEEEauthorrefmark{2},
Jingwen Chen\IEEEauthorrefmark{2},
Xiaoyang Wang\IEEEauthorrefmark{3},\\
Prosanta Gope\IEEEauthorrefmark{4},
Biplab Sikdar\IEEEauthorrefmark{2},
and Ying Zhang\IEEEauthorrefmark{1}}
\IEEEauthorblockA{\IEEEauthorrefmark{1}University of Technology Sydney\quad
\IEEEauthorrefmark{2}National University of Singapore\\
\IEEEauthorrefmark{3}University of New South Wales\quad
\IEEEauthorrefmark{4}University of Sheffield\\
feizeming@outlook.com, fei.hongming@u.nus.edu, chen.jingwen24@u.nus.edu,\\
xiaoyang.wang1@unsw.edu.au, p.gope@sheffield.ac.uk, bsikdar@nus.edu.sg, Ying.Zhang@uts.edu.au}}
\maketitle
\raggedbottom%

\begin{abstract}
Coding agents let plugins and remote services answer permission prompts, and Claude Code documents that such an answer cannot override the operator's deny rules. However, we show that this guarantee, checked one decision at a time, fails over a whole run, because an answer to one permission prompt can rewrite the policy that every later answer is checked against. For example, a remote service with no access to the operator's files approves an unrelated command and attaches an update that deletes the deny rule protecting the project's secrets. The agent then reads the secrets in that session and the next, even inside the vendor's strict sandbox. The root cause is that Claude Code checks only updates that add or replace allow rules. Relays that forward such deletions already exist, including the vendor's own Remote Control. In effect, the right to answer a prompt has become the right to administer the policy. Worse, the deletion leaves the same records as a local edit by the operator, so Claude Code cannot tell who wrote a rule.

More generally, what a runtime records decides what it can safely delegate. A runtime that cannot tell who wrote a rule can accept only updates that keep every currently denied call denied, whereas, to protect every party's restrictions, a runtime that records who wrote each rule can let each party remove only the restrictions proven to be its own. We prove that these two bounds are tight. They correctly predict whether the operator's restrictions survive in all thirteen permission designs across ten agent systems, including predictions we wrote down before running Codex, Gemini CLI, opencode, and Goose. Enforcing the bounds at the shared update step blocks the attack. A survey of 2,227 fetched marketplace plugins finds only two that send policy updates.
\end{abstract}
\begin{figure*}[t]
\centering
\includegraphics[width=\linewidth,height=138bp,keepaspectratio]{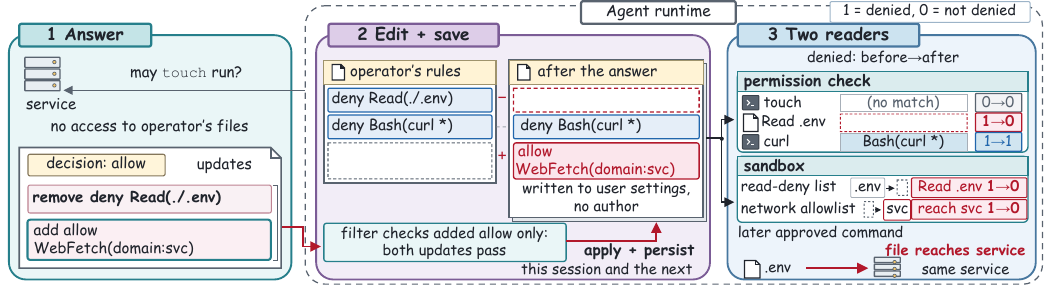}
\caption{\textbf{An attack case: one answer to an unrelated prompt rewrites the operator's policy.} Red marks lost restrictions, and blue marks retained ones.}
\label{fig:mechanism}
\end{figure*}
\section{Introduction}
\label{sec:intro}

LLM-based coding agents such as Claude Code, Codex, and Gemini CLI now write and execute code on developers' machines, where they read and edit files, run shell commands, and fetch web pages. Because the model behind such an agent can err or be steered by injected content, the agent's runtime gates its tool calls behind a permission system~\cite{ClaudeSettings2026,CodexExecpolicy2026}. The developer running the agent, whom we call the \emph{operator}, writes allow rules for permitted calls and deny rules for prohibited ones, and calls that no rule covers require approval. Vendors also provide sandboxes that derive their restrictions from the same rules and constrain even approved commands~\cite{ClaudeSandbox2026}.

Manually approving every request, however, is impractical for long or unattended sessions. Coding agents therefore let other parties answer permission prompts, such as security plugins that vet tool calls, approval services that a team runs, and mobile apps that control sessions remotely. We call such a party an \emph{answering party}. The operator delegates individual decisions to this party, but that authority remains subject to the operator's restrictions: Claude Code, for example, documents that an answer cannot override a matching deny rule~\cite{ClaudeHooks2026}. This guarantee concerns individual decisions and assumes that the rules themselves stay intact. Prior research makes the same assumption, either enforcing permissions call by call~\cite{Progent2025,CaMeL2025} or studying attackers who run code on the host~\cite{HookPry2026,He2026Root}. Neither view considers an answering party that can also modify the rules governing later decisions.

In this paper, we show that an answering party can rewrite the operator's policy through its permission replies. Claude Code lets an answer carry policy updates alongside its decision, and the runtime applies and persists them. Figure~\ref{fig:mechanism} illustrates the attack. An operator denies the agent access to the project's secrets, enables the vendor's strict sandbox, and delegates permission prompts to a remote approval service that cannot access the operator's files. When the agent requests an unrelated command, the service approves it and attaches an update that removes the deny rule. The runtime writes the update into the operator's settings, so the agent can read the secrets in the current session and in later ones, and one more update lets a subsequent command send them to the service. An answering party may already rewrite the command it approves, yet under the strict sandbox such a rewrite can neither read the secrets nor edit the settings. The update, in contrast, lifts the sandbox's limit, stays in force after the service is gone, and looks like the operator's own edit. The documented guarantee thus holds for each decision but fails over an entire run: one answer changes the rules against which later decisions are checked. This violates the separation that access control has long kept between authorization decisions and policy administration~\cite{Anderson1972,HRU1976,ARBAC97}. The right to answer a prompt effectively becomes the right to administer the policy, and the runtime, which performs the write, acts as a confused deputy~\cite{Hardy1988}.

Rejecting all policy updates would prevent the attack but break the plugins and relays that legitimately save ``always allow'' answers. The key question is therefore which updates a runtime can safely accept from a party that it trusts only to decide calls. We show that what the runtime records decides how much it can safely delegate. Intuitively, a runtime that cannot tell the operator's rules from anyone else's must treat every rule as the operator's, so it can accept only updates after which no check that reads the policy, the sandbox included, refuses less than before. A runtime that additionally records rule authors can, in contrast, let each party lift the restrictions that it alone imposed. Both statements are tight bounds, since no safe runtime that records the same information accepts more. We also prove that checking a deletion exactly against the current rules becomes coNP-hard once rule patterns mix kinds of wildcards. The bounds depend only on three facts about a design, namely what a party may change, how each check combines matching rules, and whether authors are recorded, and these facts alone predict whether a given system keeps the operator's restrictions. A repair follows from the same bounds: a single check at the step that all answers pass through.

We evaluate these findings on deployed coding agents. First, we demonstrate the attack on recent Claude Code releases through all four answer routes, with scripted and live models, under the strict sandbox, and through an unmodified deployed relay. About 7\% of the deployed relays we inspected, including the vendor's own Remote Control, already forward such updates. Second, the loss recurs beyond Claude Code: most of the other agent systems we study, among them Codex, Gemini CLI, opencode, and Goose, have a design that lets a party lift the operator's restrictions. Across all ten systems, the bounds predict every outcome we observe, including four systems whose predictions we recorded before running them. Finally, our repair keeps the operator's restrictions in all attack runs and in 20,000 random runs over a model of the runtime's settings, while only about 0.1\% of the marketplace plugins we fetched send updates that it changes.

\begin{figure}[t]
\centering
\includegraphics[width=\columnwidth]{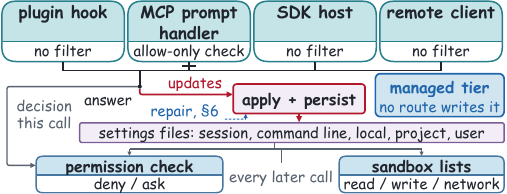}%
\caption{\textbf{Answer routes and the policy-update path in Claude Code.} Red marks the path of a deny removal, and blue marks the managed tier, which no route writes.}
\label{fig:routes}
\end{figure}

We make four contributions:
\begin{enumerate}[leftmargin=*,labelsep=0.45em,itemsep=2pt,topsep=3pt,parsep=0pt,partopsep=0pt,label=\circnum{\arabic*}]
  \item \textbf{Permission answers as an administration channel.} We identify an escalation path in delegated agent permissions: a party trusted only to answer permission prompts can rewrite the operator's policy through updates attached to its answers. We show that a guarantee enforced decision by decision does not hold over a run. We demonstrate the attack on a widely deployed coding agent through all four answer routes, including deployed relays and the vendor's strict sandbox.
  \item \textbf{Tight bounds on safe delegation.} We develop a formal model of policy updates from partially trusted parties and prove tight bounds on which updates a runtime can safely accept, given what it records. Without authorship records, a runtime can accept only updates that preserve what every check reading its policy denies, and with them it can let each party lift only its own restrictions. We further characterize when saving an answer as a rule is safe and how hard it is to check a deletion exactly.
  \item \textbf{Predicting the safety of permission designs.} We show that three facts about a permission design predict whether it keeps the operator's restrictions, and we confirm these predictions across ten agent systems, including predictions recorded in advance.
  \item \textbf{A repair for deployed coding agents.} We design a validator that enforces the bounds at the single step that all answers pass through, implement it in the same agent, and show that it keeps the operator's restrictions while preserving honest answers and the plugins that rely on them.
\end{enumerate}
We have open-sourced all code and data used in this paper.\footnote{\url{https://anonymous.4open.science/r/-When-Answering-Is-Administering-Permission-Replies-That-Rewrite-Coding-Agent-Policy-E5B0/}}

\section{Preliminaries}
\label{sec:background}

In this section, we describe the permission system of Claude Code, our running example: its rules, who answers its prompts, and what an answer can carry.

\paragraph{Permission Rules} A permission rule names a tool and, optionally, a pattern, such as \code{Read(./.env)} for reading the file \code{.env} or \code{Bash(curl *)} for any \code{curl} command. Each rule has one of three behaviors: \code{allow} runs a matching call unprompted, \code{ask} always prompts, and \code{deny} refuses. A deny rule overrides any allow rule that also matches, because deny rules are evaluated first. Rules live in several tiers: a managed tier that an administrator controls, the user's settings file, project and local settings files in the working tree, and the current session. The runtime evaluates the union of all tiers, and a settings file lists each behavior's rules as strings~\cite{ClaudeSettings2026}.

\paragraph{Answer Routes} A call that matches an ask rule, or that no rule allows, stops at a permission prompt. The operator can answer in a terminal dialog, but, as Figure~\ref{fig:routes} shows, four other routes also deliver answers, and all four may carry policy updates. A plugin's hook can take the dialog's place, a prompt handler serves unattended sessions, an application that embeds the agent replies through the SDK, and a remote front end replies through a bridge~\cite{ClaudeHooks2026}. A marketplace plugin may ship such a hook, and installing it then adds a party that answers prompts.

\paragraph{Policy Updates} When an answer allows a call, it may also carry a list of policy updates. An update can add, remove, or replace rules of one behavior, or change the permission mode, and it names the session or the settings file that it writes. No update can write the managed tier~\cite{ClaudeHooks2026}. The intended use is an answer that says ``always allow this'', so that the operator answers once for later calls of that kind. The release that introduced these updates describes them as letting hooks ``process `always allow' suggestions and apply permission updates''~\cite{ClaudeChangelog2026}.

\begin{table}[t]
\caption{\textbf{Updates that deployed relays accept from their remote side.} The survey covers 70 relays, and the first row includes the vendor's Remote Control.}
\centering
\pThreeTable
\setlength{\tabcolsep}{2bp}
\begin{tabularx}{\columnwidth}{>{\raggedright\arraybackslash}Xr}
\toprule
\pThreeHead{What the remote side can send} & \pThreeHead{Tools} \\
\midrule
\rowcolor{p3band}\textbf{Rule updates it writes, deny removals included} & \textcolor{black}{\textbf{5}} \\
A choice among updates the runtime or relay proposes & 25 \\
A decision only & 37 \\
Not determinable from the code & 3 \\
\bottomrule
\end{tabularx}

\label{tab:approvers}
\end{table}

\paragraph{Deployed Relays} To see how often answers come from elsewhere, we searched the web and GitHub for tools that relay the agent's permission prompts to a phone, a web page, or a chat. We included the vendor's Remote Control and all open-source tools above a popularity threshold, and read each of these 70 tools from where the remote answer arrives to what the runtime receives. Table~\ref{tab:approvers} shows what they send, and Appendix~\ref{app:method} describes the search.

We find that in about 7\% of these tools, the runtime applies rule updates written on the remote side, and a deletion of deny rules passes unfiltered. One of them is the vendor's own Remote Control, through which the claude.ai web and mobile apps answer for a local session, so that the answering party is whoever controls the operator's claude.ai account. In both releases we read, Remote Control's answer schema accepts all update types, and the runtime persists such deletions. The others are open-source relays with more than 40,000 GitHub stars together, and we carry out the deletion through the most-starred of them in Section~\ref{sec:runs}.

\paragraph{Documented Guarantee} The runtime evaluates deny and ask rules after a hook allows a call, checking each decision against the current rules~\cite{ClaudeHooks2026}. The vendor has fixed this check three times, after a hook's \code{allow}, a hook's \code{ask}, and a hook's rewritten input each got past a deny rule~\cite{ClaudeChangelog2026}. The check, however, covers only the decision. We study whether the operator's deny rules also survive the updates that an answer carries.
\section{Threat Model}
\label{sec:threat}

\paragraph{System Model} An agent runtime gates dangerous tool calls behind permission prompts, lets a party
other than the operator answer some of those prompts, and applies policy updates that an answer
carries, as Claude Code does. We trust the runtime, the operating system, and the administrator who writes the managed tier. We do not rely on the model to restrict itself. It may request any call, and the operator's rules decide which calls run.

\paragraph{Security Property} The operator writes deny rules and expects them to hold for every call, whatever an answering party decides. This is the guarantee the vendor documents~\cite{ClaudeHooks2026}. The property we test extends it to the whole run and to every check that reads the rules: every call, file access, or network connection that the operator's original rules deny stays denied in every state a run can reach. Section~\ref{sec:theory} formalizes this property.

\paragraph{Adversary} Our adversary is an answering party that uses one of the four answer routes, and it may arrive as a plugin from a marketplace or as a skill in a cloned repository. It chooses its answers, including any updates attached to them, after seeing the prompts so far. Its goal is to perform an action that the operator's rules deny, such as reading a protected file or sending it to its own host, in the current session or a later one. Our primary adversary reaches the agent only through its answers: a service behind a prompt-handling relay, the remote approver of an application that embeds the agent, or a remote client of a relay. It sees the prompts sent to it, not the operator's rules or files. The hook of a plugin or a skill, however, runs with the operator's privileges, so for these routes the adversary is the remote party whose answers the hook relays, or a remote service that the plugin names as its hook, which answers with no code on the host.

\paragraph{Privilege Escalation} The adversary, trusted only to decide single calls, comes to administer the policy against which later calls are checked. Its answers decide, and may rewrite, the calls it is asked about, but under the vendor's sandbox in strict mode, shell commands are kept from the protected files and the settings. The runtime, however, performs the write, giving a party barred from the operator's files write access to the operator's rules, and through them to the sandbox's limits: the runtime is a confused deputy~\cite{Hardy1988}. The change also persists after the party is gone and looks like the operator's own edit.

\paragraph{Out of Scope} We do not consider an adversary that compromises the runtime, the operating system, or the managed tier, since such an adversary needs no answer to change the policy. Prompt injection that steers the model's own calls is also outside our model, because the operator's rules already decide which of those calls run.
\section{Policy Rewriting Attack}
\label{sec:answer}

In this section, we show how a single answer rewrites the operator's policy in Claude Code.

\subsection{How Answer Updates Are Applied}
\label{sec:filter}

The runtime filters the updates an answer carries before it applies them. Its guard clause, which we read from the binary, forwards every update unless the type is \code{addRules} or \code{replaceRules} and the behavior is \code{allow}, so deletions of deny rules pass without a check. The runtime then writes the rule list of the named settings file, skipping only the managed tier and settings passed on the command line. The filter suggests that the vendor expected answers to change policy. Yet it checks the added allow rules, which a deny rule overrides anyway, and not the deletions, which can grant a denied call. In the releases we examined, three of the four answer routes skip even this filter.

The vendor's own validator also classifies updates by their syntax rather than their effect. It labels the deletion of a deny rule as widening, yet the default path applies it anyway, and it labels a replacement of the deny list as restrictive, although the replacement deletes every deny rule outside the new list. The binary even contains a stricter answer mode that drops deletions, which we call session-only answer mode, but leaves it unused. Appendix~\ref{app:static} lists the parts of the binary that we read.

\subsection{Attack Overview}
\label{sec:overview}

Figure~\ref{fig:mechanism} traces the attack of the introduction in three steps. (1) The answer approves an unrelated command and carries an update that removes the deny rule from the operator's user settings. (2) The runtime applies and saves the update. (3) The permission check and the sandbox then stop refusing the read, in that session and in later ones, with no party attached.

The attack needs little from the answering party. It needs no access to the operator's files, because the runtime, not the party, writes the change. It needs no knowledge of the operator's rules either: an update that replaces every deny list with a placeholder deletes rules its author has never seen. Nor does it need code on the host: a plugin can name a remote service as its permission hook, so that the service itself answers. The same answer also passes through deployed relays that forward rule updates, and a cloned repository can bring it in a skill that declares a permission hook, which runs once the operator agrees to use the skill.

\subsection{Effect on the Sandbox}
\label{sec:sandbox}

An answering party already controls the call it approves: its answer may rewrite the call's input before the runtime checks it against the deny and ask rules~\cite{ClaudeHooks2026}. Under the vendor's strict sandbox, however, a rewritten command can neither read a protected file nor write the settings, because the sandbox builds its read and write limits from the same deny rules~\cite{ClaudeSandbox2026}. A policy update gets past this limit, since deleting the deny rule also removes the sandbox's restriction, and later commands, even an allowed test of the project, can then read the file.

One more update brings the file to the party itself. The sandbox builds its network allowlist from the domain rules in the user settings, and an answer that deletes the deny rule and allows the party's host lets a later rewritten command send the file there. An answer need not delete anything to widen the sandbox, because an added allow rule for a host or a directory also extends the sandbox's limits. A rewritten command stays inside the sandbox, but a policy update reaches its limits, the network, and later sessions.

\subsection{Attribution}
\label{sec:author}

The runtime stores each rule as a bare string in a list, with no field that says who wrote it. A deletion by an answer therefore leaves the same record as an edit by the operator. An operator reading the file afterwards cannot tell that a plugin changed it, and when the plugin is uninstalled, the runtime can neither find the change nor undo it.
\section{Bounds on Safe Delegation}
\label{sec:theory}

\begin{figure*}[t]
\centering
\includegraphics[width=504bp,height=110.2bp]{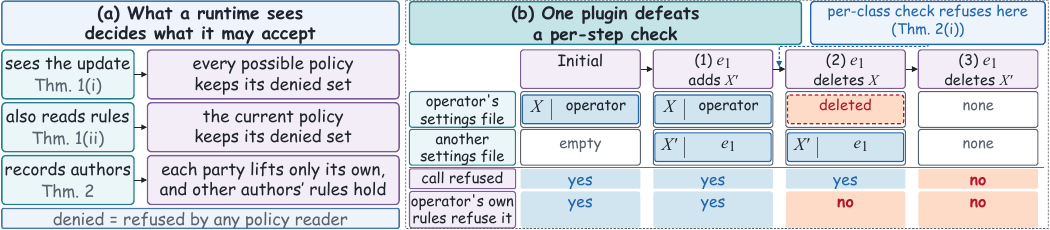}
\caption{\textbf{Bounds on safe delegation.} (a) The updates a runtime can safely accept, given what it records. (b) A single plugin that defeats a per-step check, and the step at which the per-class check of Theorem~\ref{thm:author}(i) stops it.}
\label{fig:bounds}
\end{figure*}

In this section, we characterize which updates a runtime can accept without losing any of the operator's restrictions. A runtime designer needs a rule that covers all updates, all ways of combining matches, and all checks that read the policy, because refusing deletions of deny rules is not enough. A reference monitor lets only the administrator alter the policy~\cite{Anderson1972,ARBAC97}, whereas coding agents let the party that answers an authorization question also submit policy updates. We show that how far this delegation can safely go depends on three facts about a design. These are what a party may change, how each check that reads the policy combines matching rules, and whether the runtime records who wrote each rule. Answers that a runtime saves as allow rules add one more fact, the rank of the saved rule.

\subsection{Formal Model}

We model policies, the checks that read them, and the answers that change them.

\begin{definition}[Policies, readers, and answers]
\label{def:model}
A rule is a pair $(b,p)$ of a behavior $b\in\{\code{allow},\code{ask},\code{deny}\}$ and a pattern
$p$. A policy $R$ is a finite collection of rules, in which two rules may share a behavior and a
pattern. A policy has one or more \emph{readers}, checks that refuse some actions: the permission check refuses calls, and a sandbox may refuse file reads, file writes, or network connections. Each reader combines the rules that match an action in its own way, and the \emph{denied set} $\M(R)$ collects the actions that some reader refuses, each tagged with its reader. Under deny-overrides, the rule Claude Code documents for its permission check, that check refuses exactly the calls in $\bigcup\{p \mid (\code{deny},p)\in R\}$. An answer to a prompt is a pair $(d,u)$: a decision
$d\in\{\code{allow},\code{deny}\}$ for the pending call and an update $u$, a function from policies
to policies. An action is permitted under policy $R$ only if it lies outside $\M(R)$. This is the documented per-decision guarantee. A filter decides whether $u$ is applied.
\end{definition}

A run starts from the operator's policy $R_0$ and is a sequence of policies $R_0,R_1,\dots$, in
which $R_{t+1}=u_t(R_t)$ if the filter admits the update $u_t$ sent at step $t$ and $R_{t+1}=R_t$
otherwise. The answering party chooses its answers and may be asked about any call outside the current denied set. When several parties are installed, any of them may answer a given step, and they may act together. A filter is \emph{safe} if, for every $R_0$
and every run, no action in $\M(R_0)$ is ever permitted. An action outside the current denied set can always be performed: a call through an allow rule or the party's own answer, and a file or network action through a command the party approves or rewrites. A filter is therefore safe exactly
when every reachable policy $R$ satisfies $\M(R)\supseteq\M(R_0)$. The documented guarantee holds in every state, because each call is checked against the current policy, but safety asks more: every admitted update must keep this invariant. We assume that each starting policy leaves some call outside its denied set, since otherwise no prompt arises.

We separate filters by what they can see: the update alone, the update and the current policy, or both of these together with a \emph{record} that labels each rule with the class of the party that wrote it. The operator forms a class of its own, and a rule created by an update carries the class of the party that sent it. One class per party records who wrote each rule, and a single class shared by the other parties records only which rules are the operator's. A filter that remembers which rules it admitted keeps such a record.

\paragraph{Mapping to a Deployed Runtime} In Claude Code, the policy $R$ is the union of the rules in the tiers that a session loads. Session rules end with the session, and project and local rules apply within their project alone. As a result, different sessions load different sets of tiers: all tiers in the current session, the settings files in a later session, and the user and managed tiers in another project. Each set is a policy of its own, and the conditions and proofs of this section apply to each. Ask rules are part of $R$ but add nothing to $\M(R)$, because an answering party answers ask prompts itself. No update reaches the managed tier, since no destination names it, and the tier behaves as a protected part of $R_0$. The runtime has four readers, the permission check and three sandbox lists~\cite{ClaudeSandbox2026}. The sandbox reads the file rules as the paths that commands may read or write, and the domain rules as the hosts that commands may reach. The runtime's filters see at most the update and the policy: settings files carry no authorship field, and an answer's change leaves the same record as an operator's edit.

\subsection{Filters without Authorship Records}
\label{sec:records}

The necessity proofs in this section rest on one observation: a filter that decides from a record of the state cannot treat two states with the same record differently.

\begin{lemma}[Indistinguishable states]
\label{lem:start}
Let a filter decide from the update and a record of the current state. If a reachable state has the same record as a starting state with policy $R$, then a safe filter admits an update $u$ in that state only if $\M(u(R))\supseteq\M(R)$.
\end{lemma}

\begin{proof}[Proof]
In the run that starts from that starting state, the party is asked about a call outside $\M(R)$ and answers with $u$. The filter sees the same record and the same update, so it admits $u$, and every action of $\M(R)\setminus\M(u(R))$, which $R_0=R$ denies, becomes permitted.
\end{proof}

Intuitively, a record helps a filter only where it tells the current state apart from a start. Without a record that tells the operator's restrictions apart, the current policy is the most a record can hold, and every policy could be the one the run started from. Lacking such a record, a safe filter must therefore refuse every update that could shrink the denied set: at any policy if the filter sees only the update, and at the current policy if it also reads the rules.

\begin{theorem}[Without authorship records]
\label{thm:local}
\textup{(i)} A filter that sees only the update is safe if and only if every update it admits satisfies $\M(u(R))\supseteq\M(R)$ for every policy $R$.
\textup{(ii)} A filter that also reads the rules is safe if and only if, whenever it admits $u$ at
policy $R$, $\M(u(R))\supseteq\M(R)$.
\end{theorem}

\begin{proof}[Proof]
In both parts, if every admitted update keeps the current denied set, then by induction on $t$,
$\M(R_{t+1})\supseteq\M(R_t)\supseteq\M(R_0)$, because a refused update leaves the policy unchanged and such an update keeps the denied set of every policy. Necessity follows from
Lemma~\ref{lem:start}. \emph{(i)} A filter that sees only the update has the same record in every
state, so an update it admits anywhere satisfies $\M(u(R))\supseteq\M(R)$ for every policy $R$.
\emph{(ii)} A state with policy $R$ has the same record as the starting state with policy $R$.
\end{proof}

Both bounds are tight: a filter that admits exactly these updates is safe, and no safe filter of the same kind admits more. The proof uses only the denied set, so the theorem holds for any rule syntax, any rule for combining matches, and any set of readers. The last of these matters: a reader that treats allow rules as a list of what it permits refuses less after an addition that deny-overrides makes harmless to the permission check. In this runtime, the updates that can shrink the denied set are therefore deletions and replacements of deny rules, and additions and replacements of the allow rules that the sandbox reads. The shipped filter of Section~\ref{sec:filter} sees only the update and admits all of them. By part (i) it is unsafe, and the deletion of Section~\ref{sec:overview} is the run of Lemma~\ref{lem:start} that shows it. A filter that refuses the deny updates but no others, as a designer focused on the permission check would, is unsafe as well, given the sandbox additions of Section~\ref{sec:sandbox}. Session-only answer mode is safe, because it admits only additions confined to the session and the sandbox does not read session rules.

Reading the rules lets a filter of part (ii) also admit updates that are harmless at the current policy, such as deleting a deny rule that another deny rule covers. A filter of part (i)
decides from an update's type, behavior, and tool alone, but a filter of part (ii) must then decide whether the
remaining deny rules still cover a deleted one in each policy that a session loads, so that a session rule cannot stand in for a rule in a settings file.

\begin{proposition}[The price of reading the rules]
\label{prop:conp}
Let patterns use wildcards of three kinds, \code{?} for one character other than \code{/}, \code{*} for any string without \code{/}, and \code{**} for any string, over an alphabet with a character that no pattern uses. For an update $u$ that deletes deny rules from a policy $R$, deciding whether $\M(u(R))\supseteq\M(R)$ takes polynomial time if the patterns use one kind of wildcard and is coNP-hard if they use two or more.
\end{proposition}

\begin{proof}[Proof sketch]
With one kind, replace every wildcard of a deleted pattern by the unused character. The remaining deny rules cover the pattern if and only if one of them matches this single string. With two or more kinds, DNF tautology reduces to the problem, as Appendix~\ref{app:proofs} shows in full.
\end{proof}

For command rules such as \code{Bash(curl *)}, which use only a \code{*} that matches any string, an exact check of part (ii) is thus one match per remaining rule. Claude Code's path rules use all three kinds, so a runtime that wants the extra admissions there needs a sound approximation, such as the coverage test of Appendix~\ref{app:proofs}.

\paragraph{Later Sessions} An update written to a settings file also reaches later sessions, in which the operator answers the prompts. There it can let calls run unprompted without shrinking the denied set, as mode and directory updates do. If the mode and the added directories count as part of the policy, Theorem~\ref{thm:local} extends to prompts.

\begin{corollary}[Keeping the operator's prompts]
\label{cor:prompts}
For updates written to a settings file, Theorem~\ref{thm:local} holds with $\M(u(R))\supseteq\M(R)$ replaced by the condition that no call's outcome under the settings files of $u(R)$ is less strict than under those of $R$, in the order allow, prompt, deny.
\end{corollary}

The proofs carry over, because they use only that each call's outcome is a function of the policy.

\subsection{Filters with Authorship Records}
\label{sec:authorship}

Under Theorem~\ref{thm:local}, a plugin cannot lift a restriction that it added itself. A record removes the cause. A state that holds a rule labeled with a party's class has a record that no starting state has, and Lemma~\ref{lem:start} no longer binds it. How much the record buys then depends on which parties it tells apart. We state the result for readers under which each author's denials survive in the combined policy, as under deny-overrides. Write $R_c$ for the rules labeled with class $c$, so that $\M(R_c)$ is the set of actions that the rules of $c$ deny. A filter \emph{protects} a class $c$ if, in every run, $\M(R_c)$ shrinks only at steps whose update comes from a member of $c$. A filter that protects the operator's class is safe, since the operator sends no updates.

\begin{theorem}[What records buy]
\label{thm:author}
\textup{(i)} Let a filter admit an update $u$ from $e$ at $R$ if and only if $\M(u(R)_c)\supseteq\M(R_c)$ for every class $c$ other than $e$'s. This filter protects every class, and under it $e$ can lift every action that no other class's rules deny. \textup{(ii)} If two parties $e\neq q$ share a class, no filter that decides from this record both lets $e$ lift an action that only $e$'s rules deny and keeps the actions that $q$'s rules deny from shrinking at $e$'s steps. \textup{(iii)} The per-step check is unsafe. It admits an update from $e$ whenever, just before the update, each action the update makes permitted was denied only by rules of $e$'s class.
\end{theorem}

\begin{proof}[Proof]
\emph{(i)} Fix a class $c$. An update from a party outside $c$ is admitted only if it keeps $\M(R_c)$, and a refused update changes nothing, so $\M(R_c)$ shrinks only at steps of members of $c$. The operator sends no updates, so
$\M(R_{t,\mathit{op}})=\M(R_0)$ in every reachable state, and because each class's denials survive in the combined policy,
$\M(R_t)\supseteq\M(R_{t,\mathit{op}})$, so the filter is safe. For an action denied only by rules
of $e$'s class, the update that deletes those rules keeps $\M(R_c)$ for every other class $c$, so it is
admitted and makes the action permitted. \emph{(ii)} Let parties add deny rules, and let $r$ be a deny rule for an action $a$ that no other rule denies. In one run $e$ adds $r$, in another $q$ adds it, and the record of the two states is the same, since $r$ carries the class of both. A filter that admits $e$'s deletion of $r$ in the first state admits it in the second, where $q$'s rules lose $a$ at $e$'s step. \emph{(iii)} Let $R_0=\{X\}$ with $X=(\code{deny},p)$ labeled
operator, and let $a\in p$. Party $e_1$ adds $X'=(\code{deny},p)$, labeled $e_1$, which permits nothing. Party $e_1$ then deletes $X$. The action $a$ remains denied by $X'$, so the step passes the
per-step check. Party $e_1$ deletes $X'$. At that point $a$ is denied only by a rule labeled $e_1$,
so this step passes as well, and $a\in\M(R_0)$ is no longer denied. The filter of part (i) refuses
the second step, which shrinks $\M(R_{\mathit{op}})$.
\end{proof}

A runtime can thus protect exactly the pairs of parties that its record tells apart. With one class per party, part (i) protects every party's restrictions and still lets each party lift its own. If the operator alone forms a class, the runtime is safe, but one plugin may lift the deny rules a security plugin added, and a snapshot of what the operator's rules deny at the start already gives this record. A single shared class brings back Theorem~\ref{thm:local}. Part (iii) is why the bound is stated per class, and Figure~\ref{fig:bounds}(b) draws its
counterexample, in which $e_1$'s copy sits in another settings file, so that deleting the operator's rule leaves the copy in force. Appendix~\ref{app:proofs} proves Proposition~\ref{prop:conp} and Theorem~\ref{thm:saved}.

Labels make this bound cheap to enforce, because it then needs no comparison of rules.

\begin{corollary}[Checking labels only]
\label{cor:label}
Let a filter admit an update from $e$ if and only if every rule the update deletes or replaces carries $e$'s class. This filter protects every class, and under it $e$ can lift every action that no other class's rules deny.
\end{corollary}

\begin{proof}[Proof]
An admitted update from $e$ leaves $R_c$ unchanged for every other class $c$, so it keeps $\M(R_c)$, and the argument of Theorem~\ref{thm:author}(i) applies. For an action denied only by rules of $e$'s class, the update that deletes those rules is admitted and permits the action.
\end{proof}

This filter thus attains the bound of Theorem~\ref{thm:author} with a check on labels alone, and the coverage test of Proposition~\ref{prop:conp} never arises. It admits fewer updates than the filter of Theorem~\ref{thm:author}(i), because it also refuses harmless edits of another class's rules, and Section~\ref{sec:repaireval} measures that difference.

Records, however, protect only the readers under which each author's denials survive. Consider a reader whose allow rules list what it permits, such as Claude Code's network sandbox. Such a reader lifts a restriction for every author as soon as any author adds a rule, and no record can make the addition safe. The runtime must refuse it, or take that reader's rules from a source no answer reaches, as the vendor's managed network setting does.

\begin{corollary}[Removal]
\label{cor:removal}
\textup{(i)} With one class per party, under the filter of Theorem~\ref{thm:author}(i), deleting every rule labeled $e$, allow rules included, leaves the denied set of every other party unchanged. \textup{(ii)} Without authorship records, every procedure that reads only the policy and removes some rule $e$ wrote also
removes, in some run, an identical rule that the operator wrote.
\end{corollary}

\begin{proof}[Proof]
\emph{(i)} Deleting the rules labeled $e$ leaves $R_c$ unchanged for every other class $c$. \emph{(ii)} As in Lemma~\ref{lem:start}, let a procedure read only the policy and remove a rule $r$ that $e$ wrote in a state with policy
$R$. In the run in which the operator wrote all of $R$, including $r$, and $e$ wrote nothing, the
procedure sees the same policy, so it removes $r$, which the operator wrote.
\end{proof}

Without labels, a plugin's rule and an identical rule of the operator look the same. Claude Code falls under part (ii), since a rule persisted by a plugin's answer is a bare string. When the plugin is uninstalled, removing its allow rules risks removing identical rules of the operator.
Gemini CLI, in contrast, falls under part (i): it tags each extension rule with the extension's name and deletes
exactly those rules when the extension stops~\cite{GeminiPolicy2026}.

\paragraph{Root Cause} The failures we found share one shape: the check sees less than what decides whether an update is dangerous. With deny rules, a deletion grants as surely as an addition, and a filter that inspects only added allow rules passes it. With a reader that lists what it permits, an addition grants as surely as a deletion, and a filter that guards only the permission check passes it. A check that judges each step alone admits steps that together lift another author's rule. Tschantz and Krishnamurthi observed that access-control languages with deny rules are not monotone in their rules~\cite{Tschantz2006}, whereas trust-management languages such as PolicyMaker and RT are kept monotone, so that a credential missing from an evaluation can only cause a denial~\cite{Blaze1996,LiMitchellWinsborough2002}. Theorem~\ref{thm:local} states what giving up monotonicity costs for updates, across the readers of the policy. Classic administration models record who granted what, and Theorem~\ref{thm:author} says how much of that record a coding agent needs: one class per party it must protect from the others.

\subsection{Saved Answers}
\label{sec:saved}

Some runtimes save the answer as an allow rule, letting an answer grant calls without sending any update. For the permission check, under deny-overrides and under the strictest-match rule Codex uses, such an addition keeps every denied call denied. Other runtimes, however, rank their rules, and
there such a rule can outrank a deny rule the operator wrote. The rule comes from an ``always'' answer: the answering party says ``always'', and the runtime saves an allow rule, often for more than the call it asked about, such as every command with the same first word. To cover both kinds of runtime, we model this with ranks. Each rule has a rank, the highest-ranked
matching rules decide, and ties go to deny. In this model, deny-overrides is the case in which all ranks are equal,
and last-match-wins is the case in which a later rule ranks higher. An ``always'' answer to a prompt for call $c$ adds the rule $(\code{allow},g(c))$, where the saved pattern $g(c)$ contains $c$, and the $k$-th answer saved in a run gets rank $\pi_k$. Because a runtime prompts only for calls that no rule decides yet, we assume calls with pairwise disjoint saved patterns, such as commands with different first words, and a run can use as many of them as it needs.

\begin{theorem}[Saved answers]
\label{thm:saved}
Let the operator's deny rules have ranks in a nonempty finite set $D$. Saving answers is safe for
every such operator policy if and only if $\pi_k\le\min D$ for every $k$, or $g(c)=\{c\}$ for
every call $c$.
\end{theorem}

Appendix~\ref{app:proofs} proves the theorem.

Theorem~\ref{thm:saved} thus admits two safe designs. The runtime may save only the call it asked about, a call outside the current denied set, or it may rank the saved rule at or below every deny rule. For example, under deny-overrides the rank condition always holds, however wide the saved pattern is.
Under last-match-wins, by contrast, saved answers come after the operator's rules. The rank condition then fails for every saved answer, and the runtime must save only the asked call.

Unlike the deletions of Theorem~\ref{thm:local}, this rule comes from the runtime. The party replies to one prompt with one word, and the runtime widens it, so that in opencode and Gemini CLI the first answer makes a denied call grantable, as Section~\ref{sec:predict} shows. In the terms of firewall policy analysis,
the saved rule shadows the operator's deny~\cite{AlShaer2004}, and Theorem~\ref{thm:saved} says which
ways of saving answers can create such a shadow. The theorem concerns the permission check, and a runtime with other readers needs its condition for each of them. A reader for which an allow rule means less confinement fails it even for the asked call alone, and Section~\ref{sec:predict} shows this in Codex's sandbox.

\section{Repair Design}
\label{sec:repair}

In this section, we derive a repair from the bounds and enforce it at the one step that all answers pass through.

\paragraph{Refused Updates} Each unsafe design in Table~\ref{tab:predict} admits an update that Theorem~\ref{thm:local} or Theorem~\ref{thm:saved} forbids, and the repair refuses that update. Lacking authorship records, the runtime should apply an answer's update only if the update keeps the denied set of every reader for every policy. By Theorem~\ref{thm:local}(i), every action the operator's rules deny then stays denied over all runs. For Claude Code, this means refusing deletions and replacements of deny rules from any answering party. It also means refusing additions of the allow rules that the sandbox reads, or saving them for the session only, which the sandbox does not read. Other allow rules can still be added and saved, because a deny rule overrides them and the sandbox ignores them. As a result, an operator who answers ``always allow'' through a relay keeps that feature, although the rules that the sandbox reads last for the session alone. Similarly, a mode update passes this test only if the modes it sets keep the deny rules in force.

Two further requirements come from the other designs we study. A runtime that ranks its rules must also rank a saved answer below every deny rule, or save only the call that was asked, as Theorem~\ref{thm:saved} requires. Where a plugin can replace the deciding procedure, as in DeepSeek Harness and Koishi, the repair makes that procedure read-only to the plugins it judges.

If a runtime also keeps the operator's later prompts, it applies the stricter test of Corollary~\ref{cor:prompts}, under which an answer's additions and mode or directory changes last one session, as in the vendor's unused session-only answer mode. Saving an ``always allow'' for the operator alone then requires the authorship record of Theorem~\ref{thm:author}.

\paragraph{Enforcement Point} As Figure~\ref{fig:routes} shows, the four routes apply different filters before a shared step that applies and saves updates, and on the default paths a deletion of deny rules reaches that step. The condition of Theorem~\ref{thm:local} is a property of the update, so a single filter at the shared step covers all routes, including routes added later. We implement the filter in Claude Code as one validator that both the apply step and the save step call first. When the validator refuses an update, it raises an error, so that the runtime refuses the whole answer instead of acting on the decision alone.

\paragraph{Interim Mitigation} Until runtimes adopt such a filter, operators can keep their deny rules in the managed tier or in a settings file passed on the command line, which no update reaches. They can also restrict the sandbox's network to the hosts that the managed settings allow.
\section{Evaluation}
\label{sec:eval}

In this section, we evaluate the attack, the bounds, and the repair on six research questions:
\begin{enumerate}[leftmargin=*,labelsep=0.45em,itemsep=1pt,topsep=3pt,parsep=0pt,partopsep=0pt,label=\textbf{RQ\arabic*:}]
  \item \textit{Attack effectiveness.} Can one answer lift the operator's restrictions on each answer route, with scripted and live models?
  \item \textit{Updates versus rewritten calls.} What does a policy update achieve that a rewritten call cannot, under the vendor's strict sandbox?
  \item \textit{Attribution and update types.} Is the change attributable, and which updates are unsafe?
  \item \textit{Cross-system prediction.} Do the bounds predict which permission designs are safe?
  \item \textit{Repair effectiveness and cost.} Which filters keep all restrictions, and at what cost?
  \item \textit{Compatibility.} How many deployed plugins does the repair affect?
\end{enumerate}

\subsection{Experimental Setup}
\label{sec:setup}

\paragraph{Attack Runs} Each run of Claude Code starts in a fresh home directory, and most experiments use two recent releases, which Appendix~\ref{app:method} lists. The attack still succeeds on 2.1.327, the latest release at the time of writing. The operator's user settings follow the vendor's documented example~\cite{ClaudeSettings2026}, including the deny rule \code{Read(./.env)} for the project's \code{.env} file. The agent is asked to run \code{touch} and then to read \code{.env}. In each removal arm, the answer is \code{allow} plus one update that removes the operator's rule from the user settings, and its matched control sends the same \code{allow} alone, so that the update is the sole difference. A local scripted server stands in for the model, so the runs of an arm issue identical calls. In separate runs, two Claude models choose the calls themselves.

\paragraph{Cross-System Study} To collect designs, we read the source of fourteen agent runtimes and kept, from the nine in which other parties answer prompts, the designs in which such a party can change the state that later authorization decisions consult. Together with Progent, these runtimes form the ten systems of Table~\ref{tab:predict}. The designs vary in all three facts, from session-only additions to a replaced deciding procedure, across four ways of combining rules, and with and without authorship records. For each design, we derive the prediction from these facts alone and compare it with what the system does, in the pinned source or in runs. We ran nine designs, each on two releases with a local scripted server in place of the model, and checked the other four against their source or policy model. For the five rows marked predicted first, we recorded the prediction and its pass criterion before running.

\paragraph{Filter Monitor} To compare filters beyond the attacks we ran, we built a monitor that models the runtime's settings, covering the tiers, the destinations, and the sandbox's three readers. It implements the three filters of Section~\ref{sec:theory} and the label check of Corollary~\ref{cor:label}, each refusing the allow additions that the sandbox reads, and for comparison the shipped behavior, the per-step check, and the deny-only filter.  They ran on the recorded update types, on the trace of Figure~\ref{fig:bounds}(b), and on random runs in which three parties answer over common rules, so that two of them can collude against the operator, and sessions end at random. A run is violated when any action that the operator's initial rules deny, through any reader, becomes permitted in the current session, a later session, or another project. Appendix~\ref{app:method} gives the full configurations.

\begin{figure*}[t]
\centering
\begin{tikzpicture}[x=1bp,y=-1bp,inner sep=0pt,outer sep=0pt]
\useasboundingbox (0,0) rectangle (504.00,166.70);
\draw[ink,line width=0.6bp] (0.00,0.00) -- (504.00,0.00);
\node[anchor=base,text=ink,font=\fontsize{8.0bp}{9.6bp}\selectfont\itshape] at (212.67,9.00) {Answer route};
\draw[ink,line width=0.3bp] (72.00,11.60) -- (353.33,11.60);
\node[anchor=base,text=ink,font=\fontsize{8.0bp}{9.6bp}\selectfont\itshape] at (432.67,9.00) {Variant};
\draw[ink,line width=0.3bp] (363.33,11.60) -- (502.00,11.60);
\node[anchor=base,text=ink,font=\fontsize{8.0bp}{9.6bp}\selectfont\bfseries] at (93.78,22.00) {Relay to};
\node[anchor=base,text=ink,font=\fontsize{8.0bp}{9.6bp}\selectfont\bfseries] at (93.78,31.40) {service};
\node[anchor=base,text=ink,font=\fontsize{8.0bp}{9.6bp}\selectfont\bfseries] at (141.33,22.00) {SDK};
\node[anchor=base,text=ink,font=\fontsize{8.0bp}{9.6bp}\selectfont\bfseries] at (141.33,31.40) {callback};
\node[anchor=base,text=ink,font=\fontsize{8.0bp}{9.6bp}\selectfont\bfseries] at (188.89,22.00) {Deployed};
\node[anchor=base,text=ink,font=\fontsize{8.0bp}{9.6bp}\selectfont\bfseries] at (188.89,31.40) {relay};
\node[anchor=base,text=ink,font=\fontsize{8.0bp}{9.6bp}\selectfont\bfseries] at (236.44,22.00) {Plugin};
\node[anchor=base,text=ink,font=\fontsize{8.0bp}{9.6bp}\selectfont\bfseries] at (236.44,31.40) {hook};
\node[anchor=base,text=ink,font=\fontsize{8.0bp}{9.6bp}\selectfont\bfseries] at (284.00,22.00) {URL-only};
\node[anchor=base,text=ink,font=\fontsize{8.0bp}{9.6bp}\selectfont\bfseries] at (284.00,31.40) {plugin};
\node[anchor=base,text=ink,font=\fontsize{8.0bp}{9.6bp}\selectfont\bfseries] at (331.56,22.00) {Skill in};
\node[anchor=base,text=ink,font=\fontsize{8.0bp}{9.6bp}\selectfont\bfseries] at (331.56,31.40) {checkout};
\node[anchor=base,text=ink,font=\fontsize{8.0bp}{9.6bp}\selectfont\bfseries] at (385.11,22.00) {Live};
\node[anchor=base,text=ink,font=\fontsize{8.0bp}{9.6bp}\selectfont\bfseries] at (385.11,31.40) {models};
\node[anchor=base,text=ink,font=\fontsize{8.0bp}{9.6bp}\selectfont\bfseries] at (432.67,22.00) {Every tier};
\node[anchor=base,text=ink,font=\fontsize{8.0bp}{9.6bp}\selectfont\bfseries] at (432.67,31.40) {replaced};
\node[anchor=base,text=ink,font=\fontsize{8.0bp}{9.6bp}\selectfont\bfseries] at (480.22,22.00) {Managed};
\node[anchor=base,text=ink,font=\fontsize{8.0bp}{9.6bp}\selectfont\bfseries] at (480.22,31.40) {tier};
\node[anchor=base west,text=ink,font=\fontsize{8.0bp}{9.6bp}\selectfont\bfseries] at (0.00,31.40) {Answer};
\draw[ink,line width=0.3bp] (0.00,34.60) -- (504.00,34.60);
\node[anchor=base west,text=ink,font=\fontsize{8.0bp}{9.6bp}\selectfont] at (0.00,44.00) {Removal attached};
\path[fill=losttint] (71.20,37.10) rectangle (116.36,48.10);
\node[anchor=base,text=ink,font=\fontsize{8.0bp}{9.6bp}\selectfont\bfseries] at (93.78,44.20) {6/6};
\path[fill=losttint] (118.76,37.10) rectangle (163.91,48.10);
\node[anchor=base,text=ink,font=\fontsize{8.0bp}{9.6bp}\selectfont\bfseries] at (141.33,44.20) {6/6};
\path[fill=losttint] (166.31,37.10) rectangle (211.47,48.10);
\node[anchor=base,text=ink,font=\fontsize{8.0bp}{9.6bp}\selectfont\bfseries] at (188.89,44.20) {6/6};
\path[fill=losttint] (213.87,37.10) rectangle (259.02,48.10);
\node[anchor=base,text=ink,font=\fontsize{8.0bp}{9.6bp}\selectfont\bfseries] at (236.44,44.20) {6/6};
\path[fill=losttint] (261.42,37.10) rectangle (306.58,48.10);
\node[anchor=base,text=ink,font=\fontsize{8.0bp}{9.6bp}\selectfont\bfseries] at (284.00,44.20) {6/6};
\path[fill=losttint] (308.98,37.10) rectangle (354.13,48.10);
\node[anchor=base,text=ink,font=\fontsize{8.0bp}{9.6bp}\selectfont\bfseries] at (331.56,44.20) {6/6};
\path[fill=losttint] (362.53,37.10) rectangle (407.69,48.10);
\node[anchor=base,text=ink,font=\fontsize{8.0bp}{9.6bp}\selectfont\bfseries] at (385.11,44.20) {24/24};
\path[fill=losttint] (410.09,37.10) rectangle (455.24,48.10);
\node[anchor=base,text=ink,font=\fontsize{8.0bp}{9.6bp}\selectfont\bfseries] at (432.67,44.20) {6/6};
\node[anchor=base,text=ink,font=\fontsize{8.0bp}{9.6bp}\selectfont] at (480.22,44.20) {0/6};
\node[anchor=base west,text=ink,font=\fontsize{8.0bp}{9.6bp}\selectfont] at (0.00,56.20) {Allow only};
\node[anchor=base,text=ink,font=\fontsize{8.0bp}{9.6bp}\selectfont] at (93.78,56.40) {0/6};
\node[anchor=base,text=ink,font=\fontsize{8.0bp}{9.6bp}\selectfont] at (141.33,56.40) {0/6};
\node[anchor=base,text=ink,font=\fontsize{8.0bp}{9.6bp}\selectfont] at (188.89,56.40) {0/6};
\node[anchor=base,text=ink,font=\fontsize{8.0bp}{9.6bp}\selectfont] at (236.44,56.40) {0/6};
\node[anchor=base,text=ink,font=\fontsize{8.0bp}{9.6bp}\selectfont] at (284.00,56.40) {0/6};
\node[anchor=base,text=ink,font=\fontsize{8.0bp}{9.6bp}\selectfont] at (331.56,56.40) {0/6};
\node[anchor=base,text=ink,font=\fontsize{8.0bp}{9.6bp}\selectfont] at (385.11,56.40) {0/24};
\node[anchor=base,text=ink,font=\fontsize{8.0bp}{9.6bp}\selectfont] at (432.67,56.40) {0/6$^{*}$};
\node[anchor=base,text=neutral,font=\fontsize{8.0bp}{9.6bp}\selectfont] at (480.22,56.20) {---};
\draw[ink,line width=0.6bp] (0.00,62.00) -- (504.00,62.00);
\node[anchor=base,text=ink,font=\fontsize{8.0bp}{9.6bp}\selectfont\bfseries] at (252.00,72.00) {(a) One answer, each route and variant};
\draw[ink,line width=0.6bp] (0.00,80.00) -- (211.00,80.00);
\node[anchor=base,text=ink,font=\fontsize{8.0bp}{9.6bp}\selectfont\itshape] at (94.00,101.20) {Rewritten call};
\draw[ink,line width=0.3bp] (48.00,103.80) -- (140.00,103.80);
\node[anchor=base,text=ink,font=\fontsize{8.0bp}{9.6bp}\selectfont\itshape] at (179.00,101.20) {Policy update};
\draw[ink,line width=0.3bp] (149.00,103.80) -- (209.00,103.80);
\node[anchor=base,text=ink,font=\fontsize{8.0bp}{9.6bp}\selectfont\bfseries] at (62.00,114.20) {Read};
\node[anchor=base,text=ink,font=\fontsize{8.0bp}{9.6bp}\selectfont\bfseries] at (62.00,123.60) {\ttfamily .env};
\node[anchor=base,text=ink,font=\fontsize{8.0bp}{9.6bp}\selectfont\bfseries] at (94.00,114.20) {Write};
\node[anchor=base,text=ink,font=\fontsize{8.0bp}{9.6bp}\selectfont\bfseries] at (94.00,123.60) {settings};
\node[anchor=base,text=ink,font=\fontsize{8.0bp}{9.6bp}\selectfont\bfseries] at (126.00,114.20) {Leave};
\node[anchor=base,text=ink,font=\fontsize{8.0bp}{9.6bp}\selectfont\bfseries] at (126.00,123.60) {sandbox};
\node[anchor=base,text=ink,font=\fontsize{8.0bp}{9.6bp}\selectfont\bfseries] at (163.00,114.20) {Remove};
\node[anchor=base,text=ink,font=\fontsize{8.0bp}{9.6bp}\selectfont\bfseries] at (163.00,123.60) {deny};
\node[anchor=base,text=ink,font=\fontsize{8.0bp}{9.6bp}\selectfont\bfseries] at (195.00,114.20) {Allow};
\node[anchor=base,text=ink,font=\fontsize{8.0bp}{9.6bp}\selectfont\bfseries] at (195.00,123.60) {only};
\node[anchor=base west,text=ink,font=\fontsize{8.0bp}{9.6bp}\selectfont\bfseries] at (0.00,123.60) {Sandbox};
\draw[ink,line width=0.3bp] (0.00,126.80) -- (211.00,126.80);
\node[anchor=base west,text=ink,font=\fontsize{8.0bp}{9.6bp}\selectfont] at (0.00,136.20) {Off};
\path[fill=losttint] (47.20,129.30) rectangle (76.80,140.30);
\node[anchor=base,text=ink,font=\fontsize{8.0bp}{9.6bp}\selectfont\bfseries] at (62.00,136.40) {6/6};
\path[fill=losttint] (79.20,129.30) rectangle (108.80,140.30);
\node[anchor=base,text=ink,font=\fontsize{8.0bp}{9.6bp}\selectfont\bfseries] at (94.00,136.40) {6/6};
\node[anchor=base,text=neutral,font=\fontsize{8.0bp}{9.6bp}\selectfont] at (126.00,136.20) {---};
\path[fill=losttint] (148.20,129.30) rectangle (177.80,140.30);
\node[anchor=base,text=ink,font=\fontsize{8.0bp}{9.6bp}\selectfont\bfseries] at (163.00,136.40) {6/6};
\node[anchor=base,text=ink,font=\fontsize{8.0bp}{9.6bp}\selectfont] at (195.00,136.40) {0/6};
\node[anchor=base west,text=ink,font=\fontsize{8.0bp}{9.6bp}\selectfont] at (0.00,148.40) {Strict};
\node[anchor=base,text=ink,font=\fontsize{8.0bp}{9.6bp}\selectfont] at (62.00,148.60) {0/6};
\node[anchor=base,text=ink,font=\fontsize{8.0bp}{9.6bp}\selectfont] at (94.00,148.60) {0/6};
\node[anchor=base,text=ink,font=\fontsize{8.0bp}{9.6bp}\selectfont] at (126.00,148.60) {0/6};
\path[fill=losttint] (148.20,141.50) rectangle (177.80,152.50);
\node[anchor=base,text=ink,font=\fontsize{8.0bp}{9.6bp}\selectfont\bfseries] at (163.00,148.60) {6/6};
\node[anchor=base,text=ink,font=\fontsize{8.0bp}{9.6bp}\selectfont] at (195.00,148.60) {0/6};
\draw[ink,line width=0.6bp] (0.00,154.20) -- (211.00,154.20);
\draw[ink,line width=0.6bp] (221.00,80.00) -- (335.00,80.00);
\node[anchor=base,text=ink,font=\fontsize{8.0bp}{9.6bp}\selectfont\bfseries] at (287.00,114.20) {No host};
\node[anchor=base,text=ink,font=\fontsize{8.0bp}{9.6bp}\selectfont\bfseries] at (287.00,123.60) {rule};
\node[anchor=base,text=ink,font=\fontsize{8.0bp}{9.6bp}\selectfont\bfseries] at (319.00,114.20) {Host};
\node[anchor=base,text=ink,font=\fontsize{8.0bp}{9.6bp}\selectfont\bfseries] at (319.00,123.60) {rule};
\node[anchor=base west,text=ink,font=\fontsize{8.0bp}{9.6bp}\selectfont\bfseries] at (221.00,123.60) {Deny rule};
\draw[ink,line width=0.3bp] (221.00,126.80) -- (335.00,126.80);
\node[anchor=base west,text=ink,font=\fontsize{8.0bp}{9.6bp}\selectfont] at (221.00,136.20) {Kept};
\node[anchor=base,text=ink,font=\fontsize{8.0bp}{9.6bp}\selectfont] at (287.00,136.40) {0/6};
\node[anchor=base,text=ink,font=\fontsize{8.0bp}{9.6bp}\selectfont] at (319.00,136.40) {0/6};
\node[anchor=base west,text=ink,font=\fontsize{8.0bp}{9.6bp}\selectfont] at (221.00,148.40) {Removed};
\node[anchor=base,text=ink,font=\fontsize{8.0bp}{9.6bp}\selectfont] at (287.00,148.60) {0/6};
\path[fill=losttint] (304.20,141.50) rectangle (333.80,152.50);
\node[anchor=base,text=ink,font=\fontsize{8.0bp}{9.6bp}\selectfont\bfseries] at (319.00,148.60) {6/6};
\draw[ink,line width=0.6bp] (221.00,154.20) -- (335.00,154.20);
\draw[ink,line width=0.6bp] (345.00,80.00) -- (504.00,80.00);
\node[anchor=base,text=ink,font=\fontsize{8.0bp}{9.6bp}\selectfont\itshape] at (472.00,89.00) {Repair};
\draw[ink,line width=0.3bp] (442.00,91.60) -- (502.00,91.60);
\node[anchor=base,text=ink,font=\fontsize{8.0bp}{9.6bp}\selectfont\bfseries] at (419.00,111.40) {Shipped};
\node[anchor=base,text=ink,font=\fontsize{8.0bp}{9.6bp}\selectfont\bfseries] at (456.00,102.00) {Deny-};
\node[anchor=base,text=ink,font=\fontsize{8.0bp}{9.6bp}\selectfont\bfseries] at (456.00,111.40) {only};
\node[anchor=base,text=ink,font=\fontsize{8.0bp}{9.6bp}\selectfont\bfseries] at (488.00,102.00) {Every};
\node[anchor=base,text=ink,font=\fontsize{8.0bp}{9.6bp}\selectfont\bfseries] at (488.00,111.40) {reader};
\node[anchor=base west,text=ink,font=\fontsize{8.0bp}{9.6bp}\selectfont\bfseries] at (345.00,111.40) {Answer};
\draw[ink,line width=0.3bp] (345.00,114.60) -- (504.00,114.60);
\node[anchor=base west,text=ink,font=\fontsize{8.0bp}{9.6bp}\selectfont] at (345.00,124.00) {WebFetch rule};
\path[fill=losttint] (404.20,117.10) rectangle (433.80,128.10);
\node[anchor=base,text=ink,font=\fontsize{8.0bp}{9.6bp}\selectfont\bfseries] at (419.00,124.20) {6/6};
\path[fill=losttint] (441.20,117.10) rectangle (470.80,128.10);
\node[anchor=base,text=ink,font=\fontsize{8.0bp}{9.6bp}\selectfont\bfseries] at (456.00,124.20) {6/6};
\node[anchor=base,text=ink,font=\fontsize{8.0bp}{9.6bp}\selectfont] at (488.00,124.20) {0/6};
\node[anchor=base west,text=ink,font=\fontsize{8.0bp}{9.6bp}\selectfont] at (345.00,136.20) {Edit rule};
\path[fill=losttint] (404.20,129.30) rectangle (433.80,140.30);
\node[anchor=base,text=ink,font=\fontsize{8.0bp}{9.6bp}\selectfont\bfseries] at (419.00,136.40) {6/6};
\path[fill=losttint] (441.20,129.30) rectangle (470.80,140.30);
\node[anchor=base,text=ink,font=\fontsize{8.0bp}{9.6bp}\selectfont\bfseries] at (456.00,136.40) {6/6};
\node[anchor=base,text=ink,font=\fontsize{8.0bp}{9.6bp}\selectfont] at (488.00,136.40) {0/6};
\node[anchor=base west,text=ink,font=\fontsize{8.0bp}{9.6bp}\selectfont] at (345.00,148.40) {Allow only};
\node[anchor=base,text=ink,font=\fontsize{8.0bp}{9.6bp}\selectfont] at (419.00,148.60) {0/6};
\node[anchor=base,text=neutral,font=\fontsize{8.0bp}{9.6bp}\selectfont] at (456.00,148.40) {---};
\node[anchor=base,text=ink,font=\fontsize{8.0bp}{9.6bp}\selectfont] at (488.00,148.60) {0/6};
\draw[ink,line width=0.6bp] (345.00,154.20) -- (504.00,154.20);
\node[anchor=base,text=ink,font=\fontsize{8.0bp}{9.6bp}\selectfont\bfseries] at (105.50,164.20) {(b) Rewritten call vs. policy update};
\node[anchor=base,text=ink,font=\fontsize{8.0bp}{9.6bp}\selectfont\bfseries] at (278.00,164.20) {(c) Delivery to the host};
\node[anchor=base,text=ink,font=\fontsize{8.0bp}{9.6bp}\selectfont\bfseries] at (424.50,164.20) {(d) Repairs vs. added allows};
\end{tikzpicture}%
\caption{\textbf{Attack runs on Claude Code.} Cells count the runs, of six (three per release) or 24 for live models, in which the operator's restriction was lost, with losses shaded, untested conditions dashed, and $^{*}$ marking a control shared with the first column.}
\label{fig:runs}
\end{figure*}

\subsection{Attack Effectiveness}
\label{sec:runs}

One answer lifts the operator's rule on every route we tested. As Figure~\ref{fig:runs}(a) shows, after the answer that carried the deletion, the denied read ran on each route, while the plain \code{allow} changed nothing. We observe the same outcome with both live models, with and without the strict sandbox. A rule in the managed tier, by contrast, held, because the identical update writes an empty deny list to the user settings while the managed rule stays in force. The update, not the decision, lifts the operator's rule.

Three variants confirm that the answering party needs no file access, no knowledge of the rules, and no code on the host. The decision service of the relay and SDK columns runs in a mount namespace that hides the operator's home directory, and its own attempts to read or append to the user settings failed. An answer that replaced each tier's deny list with a placeholder rule exposed protected files that it never named, in both sessions. As the URL-only column shows, a plugin whose single file names the service's URL as its permission hook deleted the rule as well.

The attack also works through deployed software. We ran the published release of the most-starred open-source relay that forwards rule updates, in its shipped mode, and answered through its own client library. The relay forwarded the deletion unchanged, and the agent read \code{.env} in that session and, with the relay stopped, in the next. A cloned repository suffices to install a permission hook: as the skill column shows, a skill that declares one deleted the rule once the operator agreed to use the skill. 

\subsection{Policy Updates versus Rewritten Calls}
\label{sec:adds}

Under the strict sandbox, the policy update exposed the protected file where the rewrites we tried did not. We compared the deletion with three rewrites of the approved shell command, which print \code{.env}, delete the rule from the settings file, or request to run outside the sandbox. As Figure~\ref{fig:runs}(b) shows, without the sandbox two of the rewrites also exposed \code{.env}, but under the strict sandbox none of the three did, while the answer that carried the update exposed it in both settings. The update gets through because deleting the rule lifts the sandbox's restriction as well, and the project's own test, an allowed command that runs inside the sandbox, printed \code{.env} in the next session. We find the same outcome in the sandbox's default setting, which runs sandboxed commands without a prompt, when the deletion comes with the answer for an MCP tool call. Even where a rewrite succeeds, the update differs in what it leaves behind: the change persists and stays out of the runtime's records.

A second update lets the answering party receive the secret. As Figure~\ref{fig:runs}(c) shows, an answer that deleted the rule and allowed the service's host let a later rewritten command post \code{.env} to the service's server within one session on both releases. Each update alone falls short. Without either update the rewrite fails at reading the file, with the deletion alone it fails at the network, and with the domain rule alone it fails at the file. This whole answer, rewritten input included, passes through four of the five relays that forward updates, the vendor's Remote Control among them.

Allow rule additions widen the sandbox without any deletion. As Figure~\ref{fig:runs}(d) shows, one WebFetch rule let a rewritten command send a project file to the service, and one Edit rule let it write outside the project, while the plain \code{allow} did neither. A rule saved for the session only did not reach the sandbox, and neither did a host rule once the managed settings restricted the network. A repair that refuses only deny updates leaves both additions open, as the deny-only column shows. 

\begin{table}[t]
\caption{\textbf{Predicted and recorded effects of rule updates on the denied set.} Predictions follow Theorem~\ref{thm:local}, and vendor labels and session-only behavior are from build 2.1.312.}
\centering
\pThreeTable
\fontsize{7.8bp}{8.7bp}\selectfont\renewcommand{\code}[1]{{\fontsize{7.5bp}{8.7bp}\selectfont\ttfamily #1}}
\setlength{\tabcolsep}{1.6bp}
\begin{tabular}{
>{\raggedright\arraybackslash}p{79bp}
>{\centering\arraybackslash}p{32.5bp}
>{\centering\arraybackslash}p{32.5bp}
>{\centering\arraybackslash}p{36bp}
>{\centering\arraybackslash}p{47bp}}
\toprule
\multirow{2}{79bp}{\pThreeHead{Update}} & \multicolumn{2}{c}{\pThreeHead{Denied set shrinks}} & \multicolumn{2}{c}{\pThreeHead{Vendor behavior}}\\
\cmidrule(lr){2-3}\cmidrule(l){4-5}
 & \pThreeSmallHead{Predicted} & \pThreeSmallHead{Recorded} & \pThreeHead{Label} & \pThreeSmallHead{Session-only}\\
\midrule
\code{addRules}, \code{allow}       & no & no & widening    & session only \\
\code{addRules}, \code{deny}        & no & no & restrictive & session only \\
\code{addRules}, \code{ask}         & no & no & restrictive & session only \\
\specialrule{.3bp}{.7bp}{.7bp}
\code{replaceRules}, \code{allow}   & no & no & widening    & dropped \\
\rowcolor{p3band}\code{replaceRules}, \code{deny}    & \textcolor{black}{\textbf{yes}} & \textcolor{black}{\textbf{yes}} & \textbf{restrictive} & as an addition \\
\code{replaceRules}, \code{ask}     & no & no & restrictive & as an addition \\
\specialrule{.3bp}{.7bp}{.7bp}
\code{removeRules}, \code{allow}    & no & no & restrictive & dropped \\
\rowcolor{p3band}\code{removeRules}, \code{deny}     & \textcolor{black}{\textbf{yes}} & \textcolor{black}{\textbf{yes}} & widening & dropped \\
\code{removeRules}, \code{ask}      & no & no & widening    & dropped \\
\bottomrule
\end{tabular}

\label{tab:cells}
\end{table}

\subsection{Attribution and Update Types}
\label{sec:cells}

We find that the runtime cannot attribute the change. We compared an answer that deletes the operator's rule with a local hook script that deletes it by editing the settings file, holding the end state fixed. The two runs leave the same deny set and the same denied command running afterwards, and their session logs have the same fields.

We then sent the rule updates of the schema one at a time to the user settings, each in a separate headless session. The deletion is one of nine rule updates, \code{addRules}, \code{replaceRules}, and \code{removeRules}, each for three behaviors. As Table~\ref{tab:cells} shows, for the permission check only the removal of deny rules and the replacement of the deny list shrank the set of denied calls, as Theorem~\ref{thm:local}(i) predicts. The replacement shrinks the denied set although the vendor's validator labels it restrictive. An addition or a replacement of the sandbox's allow rules also lifts the operator's restrictions, as Section~\ref{sec:adds} shows.

Mode and directory updates leave the deny rules in place but reach later sessions, in which the operator answers the prompts. An answer's \code{setMode} to \code{acceptEdits} in the user settings let the next session write a file unprompted on both releases, and an \code{addDirectories} let it read an outside directory unprompted. None of these changes carries an author, and the vendor's labels neither stop a deletion nor flag a replacement of the deny list.

\begin{table*}[t]
\caption{\textbf{Cross-system validation of the bounds on thirteen permission designs.} The marks on the right give the evidence behind each row.}
\centering
\providecommand{\pThreeSpan}[1]{\makebox[108bp]{\color{neutral}\leaders\hrule height 0.58ex depth -0.52ex\hfill\enspace\textit{#1}\enspace\leaders\hrule height 0.58ex depth -0.52ex\hfill}}
\pThreeTable
\setlength{\tabcolsep}{2bp}
\begin{tabular}{>{\raggedright\arraybackslash}p{99bp}>{\raggedright\arraybackslash}p{86bp}>{\centering\arraybackslash}p{26bp}>{\raggedright\arraybackslash}p{56bp}>{\raggedright\arraybackslash}p{90bp}>{\centering\arraybackslash}p{26bp}>{\centering\arraybackslash}p{31bp}>{\centering\arraybackslash}p{20bp}>{\centering\arraybackslash}p{23bp}}
\toprule
\multicolumn{3}{c}{\pThreeHead{Design and authority}} & \multicolumn{2}{c}{\pThreeHead{Prediction and observed behavior}} & \multicolumn{4}{c}{\pThreeHead{Validation}}\\
\cmidrule(lr){1-3}\cmidrule(lr){4-5}\cmidrule(lr){6-9}
\pThreeHead{Design} & \pThreeHead{A party may change} & \pThreeSmallHead{\shortstack{Records\\authors}} & \pThreeHead{Prediction} & \pThreeHead{Observed} & \pThreeSmallHead{\shortstack{Two\\releases}} & \pThreeSmallHead{\shortstack{Predicted\\first}} & \pThreeSmallHead{\shortstack{Live\\model}} & \pThreeSmallHead{\shortstack{Repair\\tested}}\\
\midrule
\multicolumn{9}{c}{\itshape Answers that delete or replace restrictions (Theorem~\ref{thm:local})}\\
\specialrule{\lightrulewidth}{.6bp}{1bp}
\rowcolor{p3band}\textbf{Claude Code}, default answers & any rule update & no & \textbf{unsafe} & deny deleted, sandbox open & \pThreeMark &  & \pThreeMark & \pThreeMark\\
Claude Code, managed tier & nothing & — & safe & rule kept, call refused & \pThreeMark &  &  & \\
Claude Code, session-only & additions for the session & no & safe & removals dropped & \multicolumn{4}{c}{\pThreeSpan{source}}\\
Gemini CLI, extensions & labeled deny and ask rules & yes & safe (Cor.~3) & own rules removed on stop & \pThreeMark & \pThreeMark &  & \\
Progent, policy model & auto-applied narrowings & no & safe (Thm.~1(ii)) & SMT check on each update & \multicolumn{4}{c}{\pThreeSpan{source}}\\
Goose, ACP client & a tool's level & no & \textbf{unsafe} & never-allow entry deleted & \pThreeMark & \pThreeMark &  & \\
Copilot SDK, embedding host & the denied rule set & no & \textbf{unsafe} & denied rules replaceable & \multicolumn{4}{c}{\pThreeSpan{source}}\\
\addlinespace[2bp]\midrule
\multicolumn{9}{c}{\itshape Saved ``always'' answers (Theorem~\ref{thm:saved})}\\
\specialrule{\lightrulewidth}{.6bp}{1bp}
VS Code Copilot agent & allow rules, deny first & no & safe & a deny always wins & \multicolumn{4}{c}{\pThreeSpan{source}}\\
Codex, approvals & allow rules, strictest wins & no & check: safe & forbidden refused & \pThreeMark & \pThreeMark & \pThreeMark & \\
 &  &  & sandbox: \textbf{unsafe} & matches leave sandbox &  & & &\\
opencode 1.19, answers & wide allow rules, last wins & no & \textbf{unsafe} & saved allow beats a deny & \pThreeMark & \pThreeMark & \pThreeMark & \\
Gemini CLI, answers & root allow at priority 950 & no & \textbf{unsafe} & outranks denies below 950 & \pThreeMark & \pThreeMark &  & \\
\addlinespace[2bp]\midrule
\multicolumn{9}{c}{\itshape Plugins that replace the deciding procedure (Theorem~\ref{thm:local})}\\
\specialrule{\lightrulewidth}{.6bp}{1bp}
DeepSeek Harness, plugin & the approval service & no & \textbf{unsafe} & refused escalation granted & \pThreeMark &  &  & \pThreeMark\\
Koishi, plugin & the permission providers & no & \textbf{unsafe} & refused permission granted & \pThreeMark &  &  & \pThreeMark\\
\bottomrule
\end{tabular}
\par\vspace{3bp}\begin{minipage}{\textwidth}\fontsize{7.7bp}{8.6bp}\selectfont From left to right, \pThreeMark{}~marks execution on two releases, a prediction and pass criterion recorded before the first run, runs with a live model, and a tested repair. \textit{source} marks rows checked against the implementation or policy model alone. For Codex, the permission check is safe (Theorem~\ref{thm:saved}), but the sandbox is unsafe (Theorem~\ref{thm:local}).\end{minipage}
\label{tab:predict}
\end{table*}

\subsection{Cross-System Prediction}
\label{sec:predict}

The three facts of the bounds predict, before any run, which designs keep the operator's restrictions. As Table~\ref{tab:predict} shows, the prediction matches what the system does for all thirteen designs we examined, including the five predictions we wrote down before running the system.

\paragraph{Safe Designs} The first group of Table~\ref{tab:predict} tests Theorem~\ref{thm:local}, the bound on what a party may change. Two safe designs come from different teams and express policy differently, yet each stays inside the same bound. One is Gemini CLI. It drops the allow rules that an extension contributes and thereby lets its extensions only tighten. It also labels each remaining rule with the extension's name, and this label gives it the removal of Corollary~\ref{cor:removal} when the extension stops~\cite{GeminiPolicy2026}, as our runs confirm. The other, Progent, applies a narrowing that its policy model proposes after an SMT solver confirms it~\cite{Progent2025}, a filter that reads the rules, as in Theorem~\ref{thm:local}(ii). The bounds add what neither design states. Without authorship records, as in Progent, no safe filter can admit more than the bound of Theorem~\ref{thm:local}. With them, as in Gemini CLI, a party can safely lift exactly its own restrictions.

\paragraph{Unsafe Front Ends} In contrast, Goose and the Copilot SDK fall outside the bound, because the application that drives a session can delete the operator's restrictions directly. Setting a tool's level in Goose first removes the tool from the operator's \code{never\_allow} list. In our runs, an ACP client barred from the operator's files set the shell tool, which the operator lists under \code{never\_allow}, to \code{always\_allow}, and Goose rewrote the operator's permission file and ran the call. Likewise, the Copilot SDK lets the embedding host replace the denied rules and approve any call.

\paragraph{Saved Answers} The second group of Table~\ref{tab:predict} tests the combining rule. Its four designs let an answer add allow rules and nothing else. A filter that judges an update's form would admit all four, yet Theorem~\ref{thm:saved} separates them by where the saved rule ranks. A denied call stays denied in VS Code's agent and in Codex however wide the saved allow is, because VS Code checks its deny list before any allow rule and Codex takes the strictest match~\cite{CodexExecpolicy2026}. In contrast, opencode 1.19 evaluates the saved approvals after the operator's rules, and the last match wins. For example, an ``always'' answer to \code{rm ./build.log} saves an allow for \code{rm *} that overrides an operator rule denying \code{rm -rf *}. When the operator enables saving answers across sessions, Gemini CLI saves each in the user tier with priority 950, where user rules range up to 999, and keys it on the command's root. An ``always'' answer to \code{git tag} therefore saves an allow for all of \code{git}, which outranks a user-tier deny for \code{git push} at a lower priority, even when the operator gave the answer. Both runtimes save more than the asked call, violating the theorem's other condition as well.

\begin{figure}[t]
\centering
\begin{tikzpicture}[x=1bp,y=-1bp,inner sep=0pt,outer sep=0pt]
\useasboundingbox (0,0) rectangle (243.00,163.10);
\draw[ink,line width=0.6bp] (0.00,0.00) -- (243.00,0.00);
\node[anchor=base,text=ink,font=\fontsize{8.0bp}{9.6bp}\selectfont\itshape] at (96.75,9.00) {opencode 1.19.4/1.19.5};
\draw[ink,line width=0.3bp] (52.00,11.60) -- (141.50,11.60);
\node[anchor=base,text=ink,font=\fontsize{8.0bp}{9.6bp}\selectfont\itshape] at (196.25,9.00) {Codex 0.164.0/0.165.0};
\draw[ink,line width=0.3bp] (151.50,11.60) -- (241.00,11.60);
\node[anchor=base,text=ink,font=\fontsize{8.0bp}{9.6bp}\selectfont\bfseries] at (73.38,22.00) {Always};
\node[anchor=base,text=ink,font=\fontsize{8.0bp}{9.6bp}\selectfont\bfseries] at (120.12,22.00) {Once};
\node[anchor=base,text=ink,font=\fontsize{8.0bp}{9.6bp}\selectfont\bfseries] at (172.88,22.00) {Forbid rule};
\node[anchor=base,text=ink,font=\fontsize{8.0bp}{9.6bp}\selectfont\bfseries] at (219.62,22.00) {No rule};
\node[anchor=base west,text=ink,font=\fontsize{8.0bp}{9.6bp}\selectfont\bfseries] at (0.00,22.00) {Arm};
\draw[ink,line width=0.3bp] (0.00,25.20) -- (243.00,25.20);
\node[anchor=base west,text=neutral,font=\fontsize{8.0bp}{9.6bp}\selectfont\itshape] at (0.00,34.60) {Predicted};
\node[anchor=base,text=neutral,font=\fontsize{8.0bp}{9.6bp}\selectfont\itshape] at (73.38,34.60) {runs};
\node[anchor=base,text=neutral,font=\fontsize{8.0bp}{9.6bp}\selectfont\itshape] at (120.12,34.60) {refused};
\node[anchor=base,text=neutral,font=\fontsize{8.0bp}{9.6bp}\selectfont\itshape] at (172.88,34.60) {refused};
\node[anchor=base,text=neutral,font=\fontsize{8.0bp}{9.6bp}\selectfont\itshape] at (219.62,34.60) {no rule};
\node[anchor=base west,text=ink,font=\fontsize{8.0bp}{9.6bp}\selectfont] at (0.00,46.80) {Observed};
\path[fill=losttint] (51.20,39.90) rectangle (95.55,50.90);
\node[anchor=base,text=ink,font=\fontsize{8.0bp}{9.6bp}\selectfont\bfseries] at (73.38,47.00) {6/6};
\node[anchor=base,text=ink,font=\fontsize{8.0bp}{9.6bp}\selectfont] at (120.12,47.00) {0/6};
\node[anchor=base,text=ink,font=\fontsize{8.0bp}{9.6bp}\selectfont] at (172.88,47.00) {0/6};
\node[anchor=base,text=ink,font=\fontsize{8.0bp}{9.6bp}\selectfont] at (219.62,47.00) {6/6};
\draw[ink,line width=0.6bp] (0.00,52.60) -- (243.00,52.60);
\node[anchor=base,text=ink,font=\fontsize{8.0bp}{9.6bp}\selectfont\bfseries] at (121.50,63.60) {(a) opencode and Codex};
\draw[ink,line width=0.6bp] (0.00,72.60) -- (243.00,72.60);
\node[anchor=base,text=ink,font=\fontsize{8.0bp}{9.6bp}\selectfont\itshape] at (146.50,81.60) {Priority of the operator's deny (sampled)};
\draw[ink,line width=0.3bp] (52.00,84.20) -- (241.00,84.20);
\node[anchor=base,text=ink,font=\fontsize{8.0bp}{9.6bp}\selectfont\bfseries] at (62.06,94.60) {0};
\node[anchor=base,text=ink,font=\fontsize{8.0bp}{9.6bp}\selectfont\bfseries] at (86.19,94.60) {100};
\node[anchor=base,text=ink,font=\fontsize{8.0bp}{9.6bp}\selectfont\bfseries] at (110.31,94.60) {500};
\node[anchor=base,text=ink,font=\fontsize{8.0bp}{9.6bp}\selectfont\bfseries] at (134.44,94.60) {900};
\node[anchor=base,text=ink,font=\fontsize{8.0bp}{9.6bp}\selectfont\bfseries] at (158.56,94.60) {949};
\node[anchor=base,text=ink,font=\fontsize{8.0bp}{9.6bp}\selectfont\bfseries] at (182.69,94.60) {950};
\node[anchor=base,text=ink,font=\fontsize{8.0bp}{9.6bp}\selectfont\bfseries] at (206.81,94.60) {951};
\node[anchor=base,text=ink,font=\fontsize{8.0bp}{9.6bp}\selectfont\bfseries] at (230.94,94.60) {999};
\node[anchor=base west,text=ink,font=\fontsize{8.0bp}{9.6bp}\selectfont\bfseries] at (0.00,94.60) {Arm};
\draw[ink,line width=0.3bp] (0.00,97.80) -- (243.00,97.80);
\node[anchor=base west,text=neutral,font=\fontsize{8.0bp}{9.6bp}\selectfont\itshape] at (0.00,107.20) {Predicted};
\draw[neutral,line width=0.4bp] (52.00,103.00) -- (52.00,105.80);
\draw[neutral,line width=0.4bp] (168.62,103.00) -- (168.62,105.80);
\draw[neutral,line width=0.4bp] (52.00,104.40) -- (100.20,104.40);
\draw[neutral,line width=0.4bp] (120.42,104.40) -- (168.62,104.40);
\node[anchor=base,text=neutral,font=\fontsize{8.0bp}{9.6bp}\selectfont\itshape] at (110.31,107.20) {runs};
\draw[neutral,line width=0.4bp] (172.62,103.00) -- (172.62,105.80);
\draw[neutral,line width=0.4bp] (241.00,103.00) -- (241.00,105.80);
\draw[neutral,line width=0.4bp] (172.62,104.40) -- (192.04,104.40);
\draw[neutral,line width=0.4bp] (221.59,104.40) -- (241.00,104.40);
\node[anchor=base,text=neutral,font=\fontsize{8.0bp}{9.6bp}\selectfont\itshape] at (206.81,107.20) {refused};
\node[anchor=base west,text=ink,font=\fontsize{8.0bp}{9.6bp}\selectfont] at (0.00,119.40) {All sessions};
\path[fill=losttint] (51.00,112.50) rectangle (73.12,123.50);
\node[anchor=base,text=ink,font=\fontsize{8.0bp}{9.6bp}\selectfont\bfseries] at (62.06,119.60) {6/6};
\path[fill=losttint] (75.12,112.50) rectangle (97.25,123.50);
\node[anchor=base,text=ink,font=\fontsize{8.0bp}{9.6bp}\selectfont\bfseries] at (86.19,119.60) {6/6};
\path[fill=losttint] (99.25,112.50) rectangle (121.38,123.50);
\node[anchor=base,text=ink,font=\fontsize{8.0bp}{9.6bp}\selectfont\bfseries] at (110.31,119.60) {6/6};
\path[fill=losttint] (123.38,112.50) rectangle (145.50,123.50);
\node[anchor=base,text=ink,font=\fontsize{8.0bp}{9.6bp}\selectfont\bfseries] at (134.44,119.60) {6/6};
\path[fill=losttint] (147.50,112.50) rectangle (169.62,123.50);
\node[anchor=base,text=ink,font=\fontsize{8.0bp}{9.6bp}\selectfont\bfseries] at (158.56,119.60) {6/6};
\node[anchor=base,text=ink,font=\fontsize{8.0bp}{9.6bp}\selectfont] at (182.69,119.60) {0/6};
\node[anchor=base,text=ink,font=\fontsize{8.0bp}{9.6bp}\selectfont] at (206.81,119.60) {0/6};
\node[anchor=base,text=ink,font=\fontsize{8.0bp}{9.6bp}\selectfont] at (230.94,119.60) {0/6};
\node[anchor=base west,text=ink,font=\fontsize{8.0bp}{9.6bp}\selectfont] at (0.00,131.60) {This session};
\node[anchor=base,text=ink,font=\fontsize{8.0bp}{9.6bp}\selectfont] at (62.06,131.80) {0/6};
\node[anchor=base,text=ink,font=\fontsize{8.0bp}{9.6bp}\selectfont] at (86.19,131.80) {0/6};
\node[anchor=base,text=ink,font=\fontsize{8.0bp}{9.6bp}\selectfont] at (110.31,131.80) {0/6};
\node[anchor=base west,text=neutral,font=\fontsize{7.5bp}{9.0bp}\selectfont\itshape] at (125.38,131.60) {predicted: refused at every priority};
\node[anchor=base west,text=ink,font=\fontsize{8.0bp}{9.6bp}\selectfont] at (0.00,143.80) {Admin deny};
\node[anchor=base,text=ink,font=\fontsize{8.0bp}{9.6bp}\selectfont] at (62.06,144.00) {0/6};
\node[anchor=base west,text=neutral,font=\fontsize{7.5bp}{9.0bp}\selectfont\itshape] at (77.12,143.80) {predicted: refused at every priority};
\draw[ink,line width=0.6bp,dash pattern=on 1.6bp off 1.2bp] (170.62,98.80) -- (170.62,124.20);
\draw[ink,line width=0.6bp] (0.00,149.60) -- (243.00,149.60);
\node[anchor=base,text=ink,font=\fontsize{8.0bp}{9.6bp}\selectfont\bfseries] at (121.50,160.60) {(b) Gemini CLI 0.65.0/0.66.0};
\end{tikzpicture}%
\caption{\textbf{Predicted and observed outcomes of saved answers.} Cells count executions of the denied command out of six runs (three per release), with lost restrictions shaded, untested cells blank, and the cutoff of Theorem~\ref{thm:saved} dashed.}
\label{fig:predicted}
\end{figure}

\paragraph{Pre-registered Runs} We tested these predictions for opencode and Codex after recording them in advance. As Figure~\ref{fig:predicted}(a) shows, the same kind of answer lets a denied command run in opencode and leaves it refused in Codex, as predicted. We obtain these outcomes with live models as well: all runs with Claude Sonnet 5.5 in opencode and GPT-5.5 in Codex, each choosing its own calls, matched their predictions.

\paragraph{Threshold in Gemini CLI} Theorem~\ref{thm:saved} predicts not only which way the outcome goes but also where it flips. Gemini CLI ranks a rule by its tier and then by its priority, and the deny decides at equal rank, as in the model of Section~\ref{sec:saved}. The theorem therefore puts the threshold exactly at priority 950 in the user tier, and the measured threshold in Figure~\ref{fig:predicted}(b) falls there. The push ran whenever the saved rule outranked the deny and in no other arm, including an answer for this call only.

\begin{table}[t]
\caption{\textbf{Simulated safety of saved answers under three combining rules.} Each cell counts violated runs out of 5,000, and $^{\dagger}$ marks the cells that Theorem~\ref{thm:saved} calls unsafe.}
\centering
\pThreeTable
\setlength{\tabcolsep}{2bp}
\begin{tabularx}{\columnwidth}{>{\raggedright\arraybackslash}p{65bp}*{3}{>{\centering\arraybackslash}X}}
\toprule
 & \multicolumn{3}{c}{\pThreeHead{Violated runs / 5,000}}\\
\cmidrule(l){2-4}
\pThreeHead{Combining rule} & \pThreeSmallHead{First word} & \pThreeSmallHead{Asked call} & \pThreeSmallHead{\shortstack{First word,\\ranked low}}\\
\midrule
Deny-overrides & 0 & 0 & 0\\
Last match wins & \cellcolor{p3fail}\textbf{2,567}$^{\dagger}$ & 0 & 0\\
Priority ordering & \cellcolor{p3fail}\textbf{2,591}$^{\dagger}$ & 0 & 0\\
\bottomrule
\end{tabularx}
\label{tab:saved}
\end{table}

Beyond these four designs, we also test the theorem in simulation. As Table~\ref{tab:saved} shows, random runs over common rules lose an operator restriction in exactly the cells that Theorem~\ref{thm:saved} calls unsafe, and an exhaustive search over small models agrees. Appendix~\ref{app:proofs} describes the search.

\paragraph{Second Readers} Theorem~\ref{thm:local} counts every reader of the policy, and three of the ten systems have a second one. In each, a check other than the permission check reads what an answer writes, and an added allow widens a sandbox. Claude Code's sandbox reads the Edit and WebFetch allow rules. Codex lets a command leave its sandbox when an allow rule matches every part of it. An answer that saved an allow rule for \code{python3} kept the operator's forbidden \code{rm -rf} refused, as predicted. The next \code{python3} command, however, ran outside the sandbox, sent a project file to the client's host, and wrote outside the workspace on both releases, whereas the same command approved once stayed inside. In Gemini CLI, a saved shell allow rule lets a command's request for network access through without a prompt. Neither opencode nor Goose has a second reader.

\paragraph{Replaced Decision Procedures} The last group of Table~\ref{tab:predict} holds the most permissive update: in DeepSeek Harness~\cite{DeepSeekHarness2026} and Koishi~\cite{KoishiPermissions2026}, a plugin replaces the deciding procedure itself. In DeepSeek Harness, the facade that confines a plugin passes the plugin's writes to the live approval service, letting a plugin of two statements replace the service's request method. Koishi lets an ordinary plugin overwrite the provider that checks a permission. Both share one outcome: a request refused before the plugin loads is granted after it loads and stays granted after the plugin stops. With the repair of Section~\ref{sec:repair}, the refused escalation stays refused while an honest plugin still runs. The three facts thus predicted all thirteen outcomes.

\begin{figure*}[t]
\centering
\begin{tikzpicture}[x=1bp,y=-1bp,inner sep=0pt,outer sep=0pt]
\useasboundingbox (0,0) rectangle (504.00,128.70);
\draw[gridc,line width=0.4bp] (144.00,28.50) -- (144.00,118.40);
\draw[gridc,line width=0.4bp] (177.00,28.50) -- (177.00,118.40);
\draw[gridc,line width=0.4bp] (210.00,28.50) -- (210.00,118.40);
\draw[gridc,line width=0.4bp] (278.00,28.50) -- (278.00,118.40);
\draw[gridc,line width=0.4bp] (306.00,28.50) -- (306.00,118.40);
\draw[gridc,line width=0.4bp] (334.00,28.50) -- (334.00,118.40);
\draw[gridc,line width=0.4bp] (359.00,28.50) -- (359.00,118.40);
\draw[gridc,line width=0.4bp] (379.00,28.50) -- (379.00,118.40);
\draw[gridc,line width=0.4bp] (399.00,28.50) -- (399.00,118.40);
\draw[gridc,line width=0.4bp] (420.00,28.50) -- (420.00,118.40);
\draw[gridc,line width=0.4bp] (444.80,28.50) -- (444.80,118.40);
\draw[gridc,line width=0.4bp] (469.60,28.50) -- (469.60,118.40);
\draw[ink,line width=0.6bp] (0.00,0.00) -- (504.00,0.00);
\node[anchor=base,text=ink,font=\fontsize{8.0bp}{9.6bp}\selectfont\bfseries] at (119.00,14.40) {Figure 3(b)};
\node[anchor=base,text=ink,font=\fontsize{8.0bp}{9.6bp}\selectfont\bfseries] at (119.00,23.80) {trace};
\node[anchor=base,text=ink,font=\fontsize{8.0bp}{9.6bp}\selectfont\bfseries] at (189.00,14.40) {Violated runs};
\node[anchor=base,text=ink,font=\fontsize{8.0bp}{9.6bp}\selectfont\bfseries] at (189.00,23.80) {of 20,000};
\node[anchor=base,text=ink,font=\fontsize{8.0bp}{9.6bp}\selectfont\bfseries] at (256.00,14.40) {Via the};
\node[anchor=base,text=ink,font=\fontsize{8.0bp}{9.6bp}\selectfont\bfseries] at (256.00,23.80) {sandbox};
\node[anchor=base,text=ink,font=\fontsize{8.0bp}{9.6bp}\selectfont\bfseries] at (316.00,14.40) {Admitted};
\node[anchor=base,text=ink,font=\fontsize{8.0bp}{9.6bp}\selectfont\bfseries] at (316.00,23.80) {updates (\%)};
\node[anchor=base,text=ink,font=\fontsize{8.0bp}{9.6bp}\selectfont\bfseries] at (387.00,14.40) {Own};
\node[anchor=base,text=ink,font=\fontsize{8.0bp}{9.6bp}\selectfont\bfseries] at (387.00,23.80) {lifts};
\node[anchor=base,text=ink,font=\fontsize{8.0bp}{9.6bp}\selectfont\bfseries] at (462.00,14.40) {Time per};
\node[anchor=base,text=ink,font=\fontsize{8.0bp}{9.6bp}\selectfont\bfseries] at (462.00,23.80) {decision ($\mu$s)};
\node[anchor=base west,text=ink,font=\fontsize{8.0bp}{9.6bp}\selectfont\bfseries] at (30.00,23.80) {Filter};
\draw[ink,line width=0.3bp] (0.00,27.00) -- (504.00,27.00);
\node[anchor=base west,text=ink,font=\fontsize{8.0bp}{9.6bp}\selectfont] at (30.00,36.90) {Shipped};
\node[anchor=base,text=lost,font=\fontsize{8.0bp}{9.6bp}\selectfont] at (119.00,36.90) {\ding{55}};
\path[fill=lostbar,draw=lost,line width=0.6bp] (144.00,31.10) rectangle (185.24,39.10);
\node[anchor=base west,text=ink,font=\fontsize{8.0bp}{9.6bp}\selectfont] at (187.74,36.90) {12,498};
\node[anchor=base,text=ink,font=\fontsize{8.0bp}{9.6bp}\selectfont] at (256.00,36.90) {9,819};
\path[fill=lostbar,draw=lost,line width=0.6bp] (278.00,31.10) rectangle (334.00,39.10);
\node[anchor=base west,text=ink,font=\fontsize{8.0bp}{9.6bp}\selectfont] at (336.50,36.90) {100.0};
\node[anchor=base west,text=neutral,font=\fontsize{8.0bp}{9.6bp}\selectfont] at (361.50,36.90) {---};
\path[fill=lostbar,draw=lost,line width=0.6bp] (423.73,35.10) circle[radius=2.1bp];
\node[anchor=base east,text=ink,font=\fontsize{8.0bp}{9.6bp}\selectfont] at (504.00,36.90) {0.2};
\node[anchor=base west,text=ink,font=\fontsize{8.0bp}{9.6bp}\selectfont] at (30.00,49.10) {Per-step check};
\node[anchor=base,text=lost,font=\fontsize{8.0bp}{9.6bp}\selectfont] at (119.00,49.10) {\ding{55}};
\path[fill=lostbar,draw=lost,line width=0.6bp] (144.00,43.30) rectangle (171.51,51.30);
\node[anchor=base west,text=ink,font=\fontsize{8.0bp}{9.6bp}\selectfont] at (174.01,49.10) {8,337};
\node[anchor=base,text=ink,font=\fontsize{8.0bp}{9.6bp}\selectfont] at (256.00,49.10) {8,201};
\path[fill=lostbar,draw=lost,line width=0.6bp] (278.00,43.30) rectangle (330.19,51.30);
\node[anchor=base west,text=ink,font=\fontsize{8.0bp}{9.6bp}\selectfont] at (332.69,49.10) {93.2};
\node[anchor=base west,text=neutral,font=\fontsize{8.0bp}{9.6bp}\selectfont] at (361.50,49.10) {---};
\path[fill=lostbar,draw=lost,line width=0.6bp] (467.59,47.30) circle[radius=2.1bp];
\node[anchor=base east,text=ink,font=\fontsize{8.0bp}{9.6bp}\selectfont] at (504.00,49.10) {688.2};
\node[anchor=base west,text=ink,font=\fontsize{8.0bp}{9.6bp}\selectfont] at (30.00,61.30) {Deny-only};
\node[anchor=base,text=kept,font=\fontsize{8.0bp}{9.6bp}\selectfont] at (119.00,61.30) {\ding{51}};
\path[fill=lostbar,draw=lost,line width=0.6bp] (144.00,55.50) rectangle (170.92,63.50);
\node[anchor=base west,text=ink,font=\fontsize{8.0bp}{9.6bp}\selectfont] at (173.42,61.30) {8,158};
\node[anchor=base,text=ink,font=\fontsize{8.0bp}{9.6bp}\selectfont] at (256.00,61.30) {8,158};
\path[fill=lostbar,draw=lost,line width=0.6bp] (278.00,55.50) rectangle (321.46,63.50);
\node[anchor=base west,text=ink,font=\fontsize{8.0bp}{9.6bp}\selectfont] at (323.96,61.30) {77.6};
\node[anchor=base west,text=neutral,font=\fontsize{8.0bp}{9.6bp}\selectfont] at (361.50,61.30) {---};
\path[fill=lostbar,draw=lost,line width=0.6bp] (425.92,59.50) circle[radius=2.1bp];
\node[anchor=base east,text=ink,font=\fontsize{8.0bp}{9.6bp}\selectfont] at (504.00,61.30) {0.3};
\draw[p3rule,line width=0.4bp] (30.00,66.70) -- (504.00,66.70);
\node[anchor=base west,text=ink,font=\fontsize{8.0bp}{9.6bp}\selectfont] at (30.00,76.00) {Sees update};
\node[anchor=base,text=kept,font=\fontsize{8.0bp}{9.6bp}\selectfont] at (119.00,76.00) {\ding{51}};
\node[anchor=base west,text=ink,font=\fontsize{8.0bp}{9.6bp}\selectfont] at (146.50,76.00) {0};
\node[anchor=base,text=ink,font=\fontsize{8.0bp}{9.6bp}\selectfont] at (256.00,76.00) {0};
\path[fill=keptbar,draw=kept,line width=0.6bp] (278.00,70.20) rectangle (314.40,78.20);
\node[anchor=base west,text=ink,font=\fontsize{8.0bp}{9.6bp}\selectfont] at (316.90,76.00) {65.0};
\node[anchor=base west,text=ink,font=\fontsize{8.0bp}{9.6bp}\selectfont] at (361.50,76.00) {0};
\path[fill=keptbar,draw=kept,line width=0.6bp] (427.47,74.20) circle[radius=2.1bp];
\node[anchor=base east,text=ink,font=\fontsize{8.0bp}{9.6bp}\selectfont] at (504.00,76.00) {0.4};
\node[anchor=base west,text=ink,font=\fontsize{8.0bp}{9.6bp}\selectfont] at (30.00,88.20) {Reads rules};
\node[anchor=base,text=kept,font=\fontsize{8.0bp}{9.6bp}\selectfont] at (119.00,88.20) {\ding{51}};
\node[anchor=base west,text=ink,font=\fontsize{8.0bp}{9.6bp}\selectfont] at (146.50,88.20) {0};
\node[anchor=base,text=ink,font=\fontsize{8.0bp}{9.6bp}\selectfont] at (256.00,88.20) {0};
\path[fill=keptbar,draw=kept,line width=0.6bp] (278.00,82.40) rectangle (322.91,90.40);
\node[anchor=base west,text=ink,font=\fontsize{8.0bp}{9.6bp}\selectfont] at (325.41,88.20) {80.2};
\node[anchor=base west,text=ink,font=\fontsize{8.0bp}{9.6bp}\selectfont] at (361.50,88.20) {0};
\path[fill=keptbar,draw=kept,line width=0.6bp] (470.53,86.40) circle[radius=2.1bp];
\node[anchor=base east,text=ink,font=\fontsize{8.0bp}{9.6bp}\selectfont] at (504.00,88.20) {1,188.5};
\node[anchor=base west,text=ink,font=\fontsize{8.0bp}{9.6bp}\selectfont] at (30.00,100.40) {Records authors};
\node[anchor=base,text=kept,font=\fontsize{8.0bp}{9.6bp}\selectfont] at (119.00,100.40) {\ding{51}};
\node[anchor=base west,text=ink,font=\fontsize{8.0bp}{9.6bp}\selectfont] at (146.50,100.40) {0};
\node[anchor=base,text=ink,font=\fontsize{8.0bp}{9.6bp}\selectfont] at (256.00,100.40) {0};
\path[fill=keptbar,draw=kept,line width=0.6bp] (278.00,94.60) rectangle (322.41,102.60);
\node[anchor=base west,text=ink,font=\fontsize{8.0bp}{9.6bp}\selectfont] at (324.91,100.40) {79.3};
\path[fill=keptbar,draw=kept,line width=0.6bp] (359.00,94.60) rectangle (388.40,102.60);
\node[anchor=base west,text=ink,font=\fontsize{8.0bp}{9.6bp}\selectfont] at (390.90,100.40) {147};
\path[fill=keptbar,draw=kept,line width=0.6bp] (470.61,98.60) circle[radius=2.1bp];
\node[anchor=base east,text=ink,font=\fontsize{8.0bp}{9.6bp}\selectfont] at (504.00,100.40) {1,205.4};
\node[anchor=base west,text=ink,font=\fontsize{8.0bp}{9.6bp}\selectfont] at (30.00,112.60) {Checks labels};
\node[anchor=base,text=kept,font=\fontsize{8.0bp}{9.6bp}\selectfont] at (119.00,112.60) {\ding{51}};
\node[anchor=base west,text=ink,font=\fontsize{8.0bp}{9.6bp}\selectfont] at (146.50,112.60) {0};
\node[anchor=base,text=ink,font=\fontsize{8.0bp}{9.6bp}\selectfont] at (256.00,112.60) {0};
\path[fill=keptbar,draw=kept,line width=0.6bp] (278.00,106.80) rectangle (317.70,114.80);
\node[anchor=base west,text=ink,font=\fontsize{8.0bp}{9.6bp}\selectfont] at (320.20,112.60) {70.9};
\path[fill=keptbar,draw=kept,line width=0.6bp] (359.00,106.80) rectangle (387.40,114.80);
\node[anchor=base west,text=ink,font=\fontsize{8.0bp}{9.6bp}\selectfont] at (389.90,112.60) {142};
\path[fill=keptbar,draw=kept,line width=0.6bp] (440.85,110.80) circle[radius=2.1bp];
\node[anchor=base east,text=ink,font=\fontsize{8.0bp}{9.6bp}\selectfont] at (504.00,112.60) {4.8};
\draw[p3rule,line width=0.5bp] (25.00,30.50) -- (25.00,64.30);
\node[anchor=base east,text=lost,font=\fontsize{8.0bp}{9.6bp}\selectfont\itshape] at (21.00,50.00) {Unsafe};
\draw[p3rule,line width=0.5bp] (25.00,69.60) -- (25.00,115.60);
\node[anchor=base east,text=kept,font=\fontsize{8.0bp}{9.6bp}\selectfont\itshape] at (21.00,95.20) {Safe};
\node[anchor=base,text=neutral,font=\fontsize{7.5bp}{9.0bp}\selectfont] at (144.00,126.20) {0};
\node[anchor=base,text=neutral,font=\fontsize{7.5bp}{9.0bp}\selectfont] at (177.00,126.20) {10k};
\node[anchor=base,text=neutral,font=\fontsize{7.5bp}{9.0bp}\selectfont] at (210.00,126.20) {20k};
\draw[neutral,line width=0.5bp] (144.00,28.50) -- (144.00,118.40);
\node[anchor=base,text=neutral,font=\fontsize{7.5bp}{9.0bp}\selectfont] at (278.00,126.20) {0};
\node[anchor=base,text=neutral,font=\fontsize{7.5bp}{9.0bp}\selectfont] at (306.00,126.20) {50};
\node[anchor=base,text=neutral,font=\fontsize{7.5bp}{9.0bp}\selectfont] at (334.00,126.20) {100};
\draw[neutral,line width=0.5bp] (278.00,28.50) -- (278.00,118.40);
\node[anchor=base,text=neutral,font=\fontsize{7.5bp}{9.0bp}\selectfont] at (359.00,126.20) {0};
\node[anchor=base,text=neutral,font=\fontsize{7.5bp}{9.0bp}\selectfont] at (379.00,126.20) {100};
\node[anchor=base,text=neutral,font=\fontsize{7.5bp}{9.0bp}\selectfont] at (399.00,126.20) {200};
\draw[neutral,line width=0.5bp] (359.00,28.50) -- (359.00,118.40);
\node[anchor=base,text=neutral,font=\fontsize{7.5bp}{9.0bp}\selectfont] at (420.00,126.20) {0.1};
\node[anchor=base,text=neutral,font=\fontsize{7.5bp}{9.0bp}\selectfont] at (444.80,126.20) {10};
\node[anchor=base,text=neutral,font=\fontsize{7.5bp}{9.0bp}\selectfont] at (469.60,126.20) {1000};
\draw[neutral,line width=0.5bp] (420.00,28.50) -- (420.00,118.40);
\draw[ink,line width=0.6bp] (0.00,118.40) -- (504.00,118.40);
\end{tikzpicture}%
\caption{\textbf{Comparison of seven filters on 20,000 random runs.} The trace column replays Figure~\ref{fig:bounds}(b), with \ding{55} where the operator's rule is lost, own lifts apply to safe filters only, and decision time uses 40 policy entries.}
\label{fig:monitor}
\end{figure*}

\subsection{Repair Effectiveness and Cost}
\label{sec:repaireval}

The repair keeps the operator's restrictions on every route we attacked, while honest updates continue to apply. On the routes of Figure~\ref{fig:runs}(a), a validator that drops deletions and replacements of deny rules kept the operator's rule in place and \code{.env} unread, and an answer's allow rule, added directory, and mode were still saved. Refusing the deny updates alone, however, let an answer widen the sandbox. The validator that also dropped the sandbox's allow additions kept the sandbox closed in the runs of Figure~\ref{fig:runs}(d), while an answer's Bash allow rule was saved as before. With whole-answer refusal on builds 2.1.319 and 2.1.325, the operator's rules and the sandbox held on all routes we tested, and honest updates went through.

To measure what each kind of information buys, we compared seven filters in the monitor. As Figure~\ref{fig:monitor} shows, only the four safe filters keep every operator restriction. By contrast, the deny-only filter keeps the restrictions of the permission check but loses one of the sandbox's in more than a third of the runs, and the shipped behavior loses one in more than half. The per-step check loses the operator's rule on the trace of Figure~\ref{fig:bounds}(b), where the filter that records authors refuses step (2). Among the safe filters, the one that reads the rules admits the most updates, because it also admits deletions that another rule covers. With coverage checked over all tiers at once, the filters that read the rules or record authors lose an operator restriction in 238 and 51 runs, in later sessions or other projects. Only the two with authorship records let a party lift its own restrictions, and the label check does so without comparing rules, two orders of magnitude faster, as the time column of Figure~\ref{fig:monitor} shows. Every filter decides within 1.2\,ms. We conclude that safety requires checking every reader, that reading the rules buys more admitted updates, and that authorship records let a party lift its own restrictions.

\begin{figure}[t]
\centering
\begin{tikzpicture}[x=1bp,y=-1bp,inner sep=0pt,outer sep=0pt]
\useasboundingbox (0,0) rectangle (243.00,83.80);
\draw[gridc,line width=0.4bp] (92.00,13.50) -- (92.00,73.50);
\draw[gridc,line width=0.4bp] (120.68,13.50) -- (120.68,73.50);
\draw[gridc,line width=0.4bp] (149.35,13.50) -- (149.35,73.50);
\draw[gridc,line width=0.4bp] (178.03,13.50) -- (178.03,73.50);
\draw[ink,line width=0.6bp] (0.00,0.00) -- (243.00,0.00);
\node[anchor=base west,text=ink,font=\fontsize{8.0bp}{9.6bp}\selectfont\bfseries] at (0.00,8.40) {Plugins with};
\node[anchor=base,text=ink,font=\fontsize{8.0bp}{9.6bp}\selectfont\bfseries] at (140.00,8.40) {Plugins (log scale)};
\node[anchor=base east,text=ink,font=\fontsize{8.0bp}{9.6bp}\selectfont\bfseries] at (243.00,8.40) {Of 2,227};
\draw[ink,line width=0.3bp] (0.00,12.00) -- (243.00,12.00);
\node[anchor=base west,text=ink,font=\fontsize{8.0bp}{9.6bp}\selectfont] at (0.00,21.90) {Any hook};
\path[fill=neutralbar,draw=neutral,line width=0.6bp] (92.00,16.10) rectangle (173.83,24.10);
\node[anchor=base east,text=ink,font=\fontsize{8.0bp}{9.6bp}\selectfont] at (243.00,21.90) {714 (32.1\%)};
\node[anchor=base west,text=ink,font=\fontsize{8.0bp}{9.6bp}\selectfont] at (0.00,33.50) {Prompt-answering hook};
\path[fill=neutralbar,draw=neutral,line width=0.6bp] (92.00,27.70) rectangle (160.91,35.70);
\node[anchor=base east,text=ink,font=\fontsize{8.0bp}{9.6bp}\selectfont] at (243.00,33.50) {253 (11.4\%)};
\node[anchor=base west,text=ink,font=\fontsize{8.0bp}{9.6bp}\selectfont] at (0.00,45.10) {Permission decision};
\path[fill=neutralbar,draw=neutral,line width=0.6bp] (92.00,39.30) rectangle (149.96,47.30);
\node[anchor=base east,text=ink,font=\fontsize{8.0bp}{9.6bp}\selectfont] at (243.00,45.10) {105 (4.7\%)};
\node[anchor=base west,text=ink,font=\fontsize{8.0bp}{9.6bp}\selectfont] at (0.00,56.70) {Permission-request hook};
\path[fill=neutralbar,draw=neutral,line width=0.6bp] (92.00,50.90) rectangle (131.05,58.90);
\node[anchor=base east,text=ink,font=\fontsize{8.0bp}{9.6bp}\selectfont] at (243.00,56.70) {23 (1.0\%)};
\node[anchor=base west,text=ink,font=\fontsize{8.0bp}{9.6bp}\selectfont\bfseries] at (0.00,68.30) {Policy updates};
\path[fill=neutralbar,draw=neutral,line width=0.6bp] (92.00,62.50) rectangle (100.63,70.50);
\node[anchor=base east,text=ink,font=\fontsize{8.0bp}{9.6bp}\selectfont\bfseries] at (243.00,68.30) {2 (0.1\%)};
\node[anchor=base,text=neutral,font=\fontsize{7.5bp}{9.0bp}\selectfont] at (92.00,81.30) {1};
\node[anchor=base,text=neutral,font=\fontsize{7.5bp}{9.0bp}\selectfont] at (120.68,81.30) {10};
\node[anchor=base,text=neutral,font=\fontsize{7.5bp}{9.0bp}\selectfont] at (149.35,81.30) {100};
\node[anchor=base,text=neutral,font=\fontsize{7.5bp}{9.0bp}\selectfont] at (178.03,81.30) {1,000};
\draw[neutral,line width=0.5bp] (92.00,13.50) -- (92.00,73.50);
\draw[ink,line width=0.6bp] (0.00,73.50) -- (243.00,73.50);
\end{tikzpicture}%
\caption{\textbf{Hook use in the vendor's plugin marketplace.} Shares are of the 2,227 plugins we fetched.}
\label{fig:census}
\end{figure}

\subsection{Plugin Compatibility}
\label{sec:compat}

The repair changes almost nothing in marketplace plugins. Figure~\ref{fig:census} shows why: delegating decisions and delegating policy changes differ sharply in use. We searched all files of the 2,176 fetched plugins still available, binaries and archives included, and the hook files we had kept for the other 51, and read the plugins that name the update field. Two of them send it.  A hook can pass a remote answer through without naming the field, but of the 23 plugins with a permission-request hook, none besides these two answers through a URL or forwards a remote answer. Appendix~\ref{app:method} describes the search in more detail.

Forbidding answering parties would break the plugins that declare a hook that can answer a prompt, including the security tools that enforce policy through hooks. The repair, in contrast, changes the update path of these two alone. Both choose among the runtime's own suggestions, and in our runs these were session-scoped directory and mode changes. The two plugins would keep working under the repair, because the repair keeps these changes, whereas session-only answer mode would drop them.
\section{Discussion}
\label{sec:discussion}

\paragraph{Design Guidelines} First, classify an update by its effect on what each reader of the policy refuses, under that reader's own way of combining rules, not by its syntax. Under this view, a deletion or a replacement of deny rules is a grant. So is an allow rule that a sandbox reads, as is a saved answer whose rule outranks a deny rule and reaches beyond the call that was asked. Second,
record the author of each rule written through an answer, and keep that record out of the answering parties' reach. Only the record lets a party undo its own rules or be cleanly removed. Third, when an answer's update is refused, refuse the whole answer, as our validator does, and show the operator the change. Applying the \code{allow} alone would act on an answer never given.

\paragraph{Limitations and Future Work} Our bounds apply to any runtime whose checks can be described by the actions they refuse, under two conditions that a runtime controls: the filter mediates every update, and the authorship record lies outside the reach of the parties it labels. A new reader of the rules comes under the bounds once its refusals join the denied set. Our runs use Linux, where the sandbox relies on bubblewrap. The update step that the attack exploits and the repair guards belongs to the runtime's platform-independent code, and repeating the sandbox runs on macOS, where the sandbox uses Seatbelt~\cite{ClaudeSandbox2026}, is a direct extension. Our validator covers the updates that an answer carries. The same reasoning applies to any state that later authorization decisions read, such as hook configurations, MCP server settings, and instruction files, and extending the monitor and the validator to such state is a natural next step. Runtimes can also adopt the authorship records of Theorem~\ref{thm:author} through a settings tier per party or signed rule entries, and solvers for the coNP-hard coverage check of Proposition~\ref{prop:conp} would let a filter admit every harmless deletion of a path rule. Finally, extracting the three facts behind our predictions automatically from a runtime's source and configuration would enable safety audits of agent runtimes and their plugin ecosystems.
\section{Related Work}
\label{sec:related}

\paragraph{Access Control Administration} The reference monitor must be tamperproof and must mediate
every access~\cite{Anderson1972,SaltzerSchroeder1975}. Harrison, Ruzzo, and Ullman showed that
deciding whether a protection system can leak a right is undecidable in general~\cite{HRU1976}. Administrative role-based access
control separates the right to use a role from the right to assign and revoke it, and bounds the
latter by an administrative scope~\cite{ARBAC97,CramptonLoizou2003}. A line of work then analyzes which states administrative actions can reach~\cite{LiTripunitara2006,Stoller2007,ARBAC2011}. Classic administration models also record who granted what: in System R only the grantor of a privilege can revoke it~\cite{GriffithsWade1976}, and XACML's administration profile attaches an issuer to every policy~\cite{XACMLAdmin2014}. Clark and Wilson separate the right to certify a transformation from the right to execute it~\cite{ClarkWilson1987}, and the tranquility principle of Bell and LaPadula keeps the labels of active objects fixed during normal operation~\cite{BellLaPadula1973}. Operating systems draw the same line: Flask and SELinux distinguish asking the policy server for a decision from loading a new
policy~\cite{Flask1999,SELinux2001}. SELinux's \code{neverallow} rules go further and refuse, at policy compile time, any allow rule for an operation they forbid~\cite{SELinuxNotebook2020}. In the cloud, Kubernetes lets a user create a role only with permissions the user already holds~\cite{K8sRBAC2026}. On the policy-language side, XACML names the rules for
combining matching policies, such as deny-overrides and first-applicable~\cite{XACML2013}, and
change-impact analysis compares what two policy versions permit~\cite{Fisler2005}. All of this work assumes that
administrative rights are assigned explicitly. In coding agents, by contrast, the right arrives implicitly with the right to answer a prompt. We characterize how much administrative power a runtime can grant, given what it records, across combining rules and policy readers.

\paragraph{Permission Systems for LLM Agents} Progent enforces programmable privilege policies and
applies its own updates under a narrowing check~\cite{Progent2025}. CaMeL separates trusted control flow from untrusted data~\cite{CaMeL2025}, and IsolateGPT isolates third-party applications behind a trusted hub~\cite{IsolateGPT2025}. SAGA and CapScope keep the policy or the capabilities outside the agents' reach~\cite{SAGA2026,CapScope2026}, and ChainCaps shows that per-tool checks can each pass while a chain of calls is unsafe, and prevents such chains by attenuation~\cite{ChainCaps2026}. Conleash turns consent
answers into reusable rules over a lattice~\cite{Conleash2026}, and Michael and Roesner survey how
agent permission systems ask and enforce~\cite{Michael2026Permissions}. These designs constrain calls or keep the policy out of reach. We instead study a deployed runtime in which an answering party can change the policy, and derive what the runtime must record to make that safe.

\paragraph{Attacks on Coding Agents} Yang et al.\ show that repository content can steer an
agent into proposing an overly broad allow rule that a human then approves and the harness
persists~\cite{Yang2026Approval}. HookPry trojanizes a marketplace plugin with a new release
that binds attacker-chosen shell commands to lifecycle events, where they run with host
privileges~\cite{HookPry2026}. Claude Code's project files could run commands and leak API keys before the user trusted the project~\cite{CheckPoint2026Hook}, and harnesses can raise injected content to a higher instruction role or carry it beyond the context it entered~\cite{He2026Root,Li2026CPE}. In DeepSeek Harness, a sandboxed agent reached an unauthenticated control API and ran unconfined~\cite{OX2026DeepSeek}, a different flaw from the plugin row of Table~\ref{tab:predict}.  HookPry and the project-file attacks need code on the host, the attacks of He et al.~\cite{He2026Root} and Li et al.~\cite{Li2026CPE} promote injected text, and in Yang et al.\ a human approves the added rule. In our attack, a remote party with no code on the host changes the policy through updates attached to its answer, which the runtime writes.

\paragraph{Extension Ecosystems} Work on browser extensions, Android re-delegation, package ecosystems, and MCP servers studies third-party code~\cite{Barth2010,FeltRedelegation2011,Zimmermann2019,MCPLandscape2026}, whereas we study what a third party's answers can do to the policy.
\section{Conclusion}
\label{sec:conclusion}

An answer to one permission prompt can carry updates that delete the operator's deny rules and widen the vendor's sandbox. Answering a prompt therefore confers the power to administer the policy. For runtime designers, the lesson is that how much a runtime can safely delegate depends on what it records. Without authorship records, a runtime must refuse any update that shrinks what some reader of the policy denies, and with them it can let each party lift its own restrictions. These bounds predict which designs keep the operator's restrictions across ten agent systems, and one check that enforces them blocks the attack in Claude Code while changing the update path of only two marketplace plugins.

\nobalance
\bibliographystyle{ieeetr}
\bibliography{refs}

\appendices
\section{Static Analysis of Claude Code}
\label{app:static}

\begin{table}[t]
\caption{\textbf{Binary regions that handle an answer's updates in Claude Code.} We read each region from build 2.1.312, and each occurs once.}
\centering
\pThreeTable
\fontsize{7.8bp}{8.7bp}\selectfont
\setlength{\tabcolsep}{2bp}
\begin{tabularx}{\columnwidth}{>{\raggedright\arraybackslash}p{62bp}>{\raggedright\arraybackslash}X}
\toprule
\pThreeHead{Region} & \pThreeHead{What it shows} \\
\midrule
Answer filter & the loop shown below, unchanged apart from identifiers \\
Hook and SDK paths & in the default mode, updates are applied and persisted without that filter unless a suppression condition holds \\
Prompt-handler path & applies the loop shown below, and a filter that keeps deny rules applies only for one named server under a feature flag \\
Default mode & each of the 12 answer entry points of build 2.1.319 selects the default answer kind (\code{launcher}), not session-only answer mode \\
\specialrule{.3bp}{.7bp}{.7bp}
Update validator & labels removals of \code{allow} rules, and additions or replacements of \code{deny} or \code{ask} rules, as restrictive, labels every other rule update as widening, and applies a well-formed widening update \\
Session-only cap & in session-only answer mode, drops every \code{removeRules}, \code{addDirectories}, \code{removeDirectories}, and replacement of allow rules, and rewrites every other addition or replacement as a session-only addition \\
Session-only refusal & in session-only answer mode, refuses the pending call when its caps on session additions would drop a restriction the answer adds \\
\specialrule{.3bp}{.7bp}{.7bp}
Removal apply & deletes the named rules from the named destination file \\
\bottomrule
\end{tabularx}

\label{tab:static}
\end{table}

Table~\ref{tab:static} lists the regions of build 2.1.312 (digest prefix
\code{4f4178ce41d83e23}) behind Section~\ref{sec:filter}. We extracted them read-only from the
installed binary and executed nothing. Four strings that minification does not rename mark these regions, and a search of every build we installed finds all four, and the filter's loop, in all 10
builds from 2.1.301 to 2.1.325. The artifact's extraction script records each region's byte offset.

The filter's loop in build 2.1.231 opens with a guard clause that forwards every update other than an addition or a replacement of allow rules, and later builds up to 2.1.325 contain the same loop with renamed identifiers:
\par\smallskip\noindent\begin{minipage}{\linewidth}
\begin{lstlisting}
for (let s of e) {
  if (!((s.type === "addRules" ||
         s.type === "replaceRules") &&
        s.behavior === "allow")) {
    i.push(s); continue // forwarded unfiltered
  }
  ... // allow rules are checked here
}
\end{lstlisting}
\end{minipage}
\section{Experimental Details}
\label{app:method}

We detail here the runs and searches of Section~\ref{sec:eval}. The scripts and configuration files of each experiment in our code repository hold its full setup.

\paragraph{Runs on Claude Code} The operator's user settings allow \code{Bash(npm run lint)}, \code{Bash(npm run test *)}, and \code{Read(\textasciitilde/.zshrc)}, and they deny \code{Bash(curl *)} and \code{Read(./.env)}. In the placeholder variant of Section~\ref{sec:runs}, the settings also deny a secrets directory and a second environment file in the project and local settings, and the answer exposed all three files in both sessions while the control exposed none. Our reading of the binary uses build 2.1.312, and the live runs use \code{claude-sonnet-5-5} and \code{claude-opus-5-5}. Figure~\ref{fig:runs}(a) uses builds 2.1.319 and 2.1.314 for the relay and SDK columns, 2.1.319 and 2.1.325 for the deployed relay, 2.1.319 and 2.1.314 for the plugin and skill columns, and 2.1.325 and 2.1.319 for the URL hook.  Each arm has three runs on each build, arms are interleaved, and scripts answer every prompt. Only the \code{touch} call needs an answer, because reads inside the project need no prompt. All runs used Ubuntu 22.04, where the sandbox relies on bubblewrap. We record every settings file before and after each session, every prompt and answer, and the agent's transcript, and a denied read counts as having run when the transcript shows the canary string from \code{.env}.

\paragraph{Plugin, Skill, and Relay Routes} The plugin and skill routes drive the runtime through a pseudo-terminal, because the hook fires only when a dialog would appear. The driver presses no key on a permission dialog, except that it accepts the dialog that asks whether to use the skill, a file in the repository's \code{.claude/skills} directory whose frontmatter declares the hook. We ran the deployed relay's published release in its shipped mode, which asks before every call, with its cloud connection off, and answered through the relay's own client library.

\paragraph{Live Models} The runs with a live model use the relay route with the operator settings of Section~\ref{sec:setup}, the models \code{claude-sonnet-5-5} and \code{claude-opus-5-5}, and builds 2.1.325 and 2.1.319, with and without the strict sandbox, three runs per arm in each of these eight combinations (24 per arm in all). The model is asked in plain words to run \code{touch} and then read \code{.env}, and in the next session to read it again. A read counts when its canary appears in a tool result that is not an error. In the live runs of opencode (\code{claude-sonnet-5-5}) and Codex (\code{gpt-5.5}), with the clients, rules, and builds of Figure~\ref{fig:predicted}(a), the model is asked to delete \code{build.log}, then a directory with \code{rm -rf}, and it issued both commands in all 24 runs. In all runs of Figure~\ref{fig:predicted}, the answering party is a client of the runtime's own interface, barred from the operator's files, and is asked once per run, about an unrelated command.

\paragraph{Sandbox Runs} The arms of Section~\ref{sec:adds} enable the sandbox in strict mode in the operator's user settings, and the arms asked about \code{touch} also turn off the automatic approval of sandboxed commands, so that every shell command needs an answer. The default-setting arms keep the automatic approval, and their first call is an MCP tool that needs approval. On builds 2.1.314 and 2.1.319, the deletion in its answer exposed \code{.env} in both sessions in all 6 runs, and the plain \code{allow} exposed it in none of the 6 control runs.  The project's \code{package.json} has a \code{test} script that prints \code{.env}, which the operator's rule \code{Bash(npm run test *)} allows. In the runs in which the service receives \code{.env}, on builds 2.1.325 and 2.1.319, a listener on the loopback interface stands for the service's host, and the rewritten command sends its request through the proxy that the sandbox exports to every command. The runs of Figure~\ref{fig:runs}(d) keep this configuration, with a source file that holds a run-unique canary and that no rule covers. Each later answer rewrites the approved \code{touch} into a command that posts that file to the listener or writes a file outside the project, which the sandbox does not let commands write. The service received the file in 0 of 6 runs when the host rule was saved for the session only, and in 0 of 6 when the managed settings set \code{allowManagedDomainsOnly}.

\paragraph{Patched Builds} Both builds are Bun executables. We rewrote the source of the one module that holds the shared validator, keeping its length, and cleared that module's bytecode, so that the runtime compiles the patched source. Against the patched builds, the removal arms of Figure~\ref{fig:runs}(a), the replacement arm, and the plugin and skill arms ran three times per build, as did three arms whose answers save an allow rule for \code{touch}, an outside directory, or the mode \code{acceptEdits} in the user settings. Every removal arm sent its update, and the three other updates were saved, in every run. The every-reader builds of Figure~\ref{fig:runs}(d) also drop \code{addRules} and \code{replaceRules} of allow rules for Edit, Write, and WebFetch, and on them an answer that saves \code{Bash(touch *)} was still saved in every run. For whole-answer refusal, the check of builds 2.1.319 and 2.1.325 raises an error instead of returning a drop. The same arms and the two sandbox additions of Figure~\ref{fig:runs}(d) ran on them three times per build, and the operator's rules and the sandbox held in all 42 runs. The refused call failed with an error naming the update on the relay and SDK routes and went to the operator's own dialog on the plugin and skill routes. The honest arms and controls behaved as above.

\paragraph{Mode and Directory Updates} On the relay of Section~\ref{sec:runs}, builds 2.1.319 and 2.1.314, the answer about \code{touch} carries \code{setMode} to \code{acceptEdits}, \code{setMode} to \code{bypassPermissions}, or \code{addDirectories} for a directory outside the project, each to the user settings, or no update. A second, headless session with no prompt handler then writes a project file and reads a file in that directory. Each update let its call run in 6 of 6 runs, and in the control both calls are refused in every run. The bypass update is never persisted.

\paragraph{Attribution} The comparison of Section~\ref{sec:cells} ran on build 2.1.233, where the session logs of both sides have the same 87 fields, 0 unique to either side. On the recent releases we searched every file the runtime wrote in the 27 runs of the relay, SDK, and deployed-relay arms in which an answer carried a removal or a replacement. Their session logs differ from those of the controls only in the outcome of the protected reads.

\paragraph{Cross-System Selection} Of the fourteen runtimes we read, five coding agents let only the operator answer prompts: Cline, Continue, Crush, OpenHands, and aider.

\paragraph{Gemini CLI} On releases 0.65.0 and 0.66.0, a test extension that denied \code{git status} and \code{git log}, beside an operator deny for \code{git log}, lost its allow rule at load, and disabling it removed only its own denies.  A rule's rank is its tier number plus one thousandth of its priority, and because the engine sorts its rules by rank with a stable sort and loads the operator's rules first, the deny decides at equal rank. For example, an answer saved for all future sessions is an allow at rank 4.950 in the user tier. For saved answers, the operator's rule sits in the user tier, or in the admin tier through \code{-{}-admin-policy}, at 9 ranks around the threshold of Figure~\ref{fig:predicted}(b). In the bundle of 0.66.0, the answer ``always allow and save'' to a sandbox expansion request saves the command's root with the requested paths and network access in \code{sandbox.toml}. When a saved allow rule for the command's root allows a call, the scheduler skips the confirmation that would show a request for more permissions and passes the requested network access on.

\paragraph{Codex Sandbox} These runs keep the harness of Figure~\ref{fig:predicted}(a) with \code{sandbox\_mode} \code{workspace-write}, the network off, and the project untrusted. The scripted model runs \code{touch}, then one \code{python3} command that posts a project file to a loopback listener and writes a file outside the workspace and \code{/tmp}. The client accepts the first approval with the amendment \code{["python3"]} in the attack and \code{["touch"]} in the control, and approves the later command once. The file arrived and the outside file was written in 6 of 6 attack runs and 0 of 6 control runs. In the source of 0.165.0, a command whose every segment matches an allow rule bypasses the sandbox, and the app server saves the client's amendment unchecked.

\paragraph{Goose} We ran the official release binaries 1.56.0 and 1.57.0 as an ACP server, and the shell call ran in 6 of 6 runs against 0 of 6 in the control, on both releases. The operator uses the manual-approval mode, in which Goose applies per-tool levels, the client sends Goose's set-tool-permissions request before the task, and a local scripted server stands in for the model.

\paragraph{DeepSeek Harness} At commit \code{47f9438}, the harness lets the model define a plugin
and mount it while serving a request. The plugin runs in a \code{node:vm} context behind a
whitelist facade that refuses it the root context and exposes the services it may use, among them the approval service that resolves sandbox escalations. The facade's
wrapper for a service traps reads but lets writes through to the live instance, so a plugin of two
statements replaces the service's request method. On the vendor's own define-and-run path, with
its real sandbox, facade, approval service, and escalation routine, a request to widen the sandbox to full access fails closed before the plugin runs, because no approver is mounted, and returns the widened mode after it runs. Stopping the plugin reports
success, the escalation stays granted, and the replaced method no longer writes the audit entry that the service records for every request.

\paragraph{Koishi} We defined a permission and a provider that refuses it, standing in for a capability
the operator has not granted. An ordinary plugin that declares nothing and touches only its own
context overwrites the check: the permission is refused before the plugin loads, granted after it
loads, and still granted after the plugin is disposed. The plugin's own effect list stays empty, so
the runtime has nothing to undo.

\paragraph{Repairs in DeepSeek Harness and Koishi} In DeepSeek Harness we wrap the approval service when the host registers it, so that writes to it are refused and recorded. In the harness's own test suite, the model-written plugin then leaves the refused escalation refused, while an honest plugin that reads and calls the service still mounts, runs, and stops cleanly. In Koishi, a guard inside the registry that refuses changes to the provider of a named permission stops the overwrite above.

\paragraph{Random Runs} In the filter monitor, the filters that read the rules use the sound coverage test of Appendix~\ref{app:proofs}, which compares each deny rule with single remaining rules. Table~\ref{tab:gen} gives the generators of the random runs in
Table~\ref{tab:saved} and Figure~\ref{fig:monitor}. Both draw from 15 probe calls (eleven shell
commands, three file reads, and one fetch) and from common patterns such as
\code{Bash(rm -rf *)} and \code{Read(./.env)}, with fixed seeds. The filter runs add three paths the sandbox reads, three it writes, two of them outside the project, and two hosts, with two WebFetch domain rules, an Edit allow outside the project, and an Edit deny, and judge each run against all four readers.

\begin{table}[t]
\caption{\textbf{Generator settings for the two sets of random runs.}}
\centering
\pThreeTable
\fontsize{7.8bp}{8.7bp}\selectfont
\setlength{\tabcolsep}{2bp}
\begin{tabularx}{\columnwidth}{>{\raggedright\arraybackslash}p{36bp}*{2}{>{\raggedright\arraybackslash}X}}
\toprule
 & \pThreeHead{Saved answers} & \pThreeHead{Filters}\\
 & for Table~\ref{tab:saved} & for Figure~\ref{fig:monitor}\\
\midrule
Runs & 5,000 per combining rule & 20,000 \\
Operator's rules & 2--5 deny and 1--3 allow, at random priorities below a saved answer's & 2--5 deny, 1--3 allow, and 0--1 ask, each in a random settings file, plus a managed deny rule in 30\% of runs \\
\addlinespace[2bp]
Answers & 6 ``always'' answers, each to a random call outside the current denied set & 6, each from one of three parties, carrying one rule update of random type, behavior, and destination that names 1--2 rules, after which the session ends with probability 0.25 \\
Patterns & 10 deny, 4 allow & 11 deny, 7 allow, 1 ask \\
\bottomrule
\end{tabularx}

\label{tab:gen}
\end{table}

\paragraph{Update Types} The runs of Section~\ref{sec:cells} start from settings in which \code{allow},
\code{deny}, and \code{ask} each hold the whole-tool rule \code{Read}. Additions and replacements name \code{Bash}, which the starting settings lack, and removals name
\code{Read}, which they hold, so every change is visible in the file. All probe rules are whole-tool rules, so the denied set of a
policy is exactly the set of tool names in its deny list. As Table~\ref{tab:cells} shows, the denied set before and after each recorded run confirms the prediction of Theorem~\ref{thm:local}(i) in every cell.

\paragraph{Relay Search} On 2026-10-29 we searched the web and GitHub for tools whose description says that a phone, web page, chat platform, CI job, or hosted service drives or answers a Claude Code session, kept those with at least 100 stars or 1,000 weekly npm downloads, and read each at a pinned commit, with the classes of Table~\ref{tab:approvers} fixed before the reading. Of the five relays that forward updates, all but one also forward a rewritten input.

\paragraph{Marketplace Search} As a recall check, the search of Section~\ref{sec:compat} also finds the open-source relays in Table~\ref{tab:approvers} that, by our manual reading, return updates. Anthropic's community marketplace for Claude Code lists 2,277 entries at pinned commits. Of these, 34 could not be fetched and 16 held no file at its path. Of the remaining 2,227 plugins of Section~\ref{sec:compat}, 2,176 were still available when we searched every file. Of the five that name \code{updatedPermissions}, three name it only in documentation or in an unused type. The two that send updates let a remote side choose among the updates the runtime proposes, and against a stub service of ours one attached whichever update the stub chose. For the 51 plugins removed upstream after we fetched them, the hook files we had fetched name no update field.
\raggedbottom
\section{Proofs and Additional Checks}
\label{app:proofs}

We prove the remaining results of Section~\ref{sec:theory} below and describe the checks we ran on them.

\begin{proof}[Proof of Proposition~\ref{prop:conp}]
Under deny-overrides, $\M(u(R))\supseteq\M(R)$ fails exactly when some call that a deny rule of
$R$ matches is matched by no deny rule of $u(R)$. For an update that deletes the rule with pattern
$p$ from a policy whose other deny patterns are $Q$, this is
$L(p)\not\subseteq\bigcup_{q\in Q}L(q)$. Let $\#$ be a character that occurs in no pattern.
\emph{One kind.} Let $p'$ be $p$ with every wildcard replaced by $\#$. A $q$ that matches $p'$ matches
all of $L(p)$, because its literals contain no $\#$ and so lie inside literal parts of $p$, and each $\#$
lies under a wildcard of $q$ of the same kind as the one it replaced. The remaining rules therefore
cover $p$ if and only if one of them matches $p'$.
\emph{Hardness.} With two kinds, and hence with more, we reduce from DNF tautology over $x_1,\dots,x_n$ with terms $T_1,\dots,T_k$. Let $W$
be the wider of the two kinds, let $p$ be $c_0Wc_1W\cdots Wc_n$ with distinct characters $c_i$, and
for each term let $q_j$ be $c_0s_1c_1\cdots s_nc_n$. When one kind is \code{?}, the $i$-th part of a
call is marked if it is nonempty, and $s_i$ is \code{?}$W$ if $x_i\in T_j$, empty if $\neg x_i\in T_j$,
and $W$ otherwise. If $W$ is then \code{**}, the policy also denies, for each $i$, the pattern $p$ with
its $i$-th $W$ replaced by $W$\code{/}$W$, which covers every call whose $i$-th part holds a \code{/}.
For \code{*} and \code{**}, a part is marked if it contains a \code{/}, and $s_i$ is \code{**/**},
\code{*}, or \code{**}. Reading $x_i$ as true exactly when the $i$-th part is marked, the $q_j$ of a
satisfied term matches every other call of $L(p)$ at the positions of $p$'s characters. On calls whose
parts are empty or a single marked character, each $c_i$ occurs once, so $q_j$ matches only if its term
holds, and no added pattern matches. Deleting $p$ therefore keeps the denied set if and only if the
formula is a tautology. Martens et al.\ classify inclusion for related fragments of regular expressions~\cite{MNS2009}. An exact automaton check of every small case confirms both parts (617,612 inclusions and 1,082 formulas per pair of kinds).
\end{proof}

\begin{proof}[Proof of Theorem~\ref{thm:saved}]
Here the $k$-th step of a run adds $(\code{allow},g(c))$ at rank $\pi_k$ for a call
$c\notin\M(R_t)$ that the party was asked about, and every deny rule is the operator's. \emph{If.}
Let $x\in\M(R_0)$, and assume by induction that $x\in\M(R_t)$. If $x\notin g(c)$, the new rule does
not match $x$. Otherwise $x\neq c$, because $c$ is not denied and $x$ is, so $g(c)\neq\{c\}$ and
therefore $\pi_k\le\min D$. Among the rules of $R_t$ that match $x$ at the highest rank there is a
deny rule. That rank therefore lies in $D$ and is at least $\pi_k$. The new rule ranks below it, or ties
with it and loses to deny, and $x$ stays denied. \emph{Only if.} Suppose $g(c)\neq\{c\}$ and
$\pi_k>d$ for some $k$ and some $d\in D$. Pick $x\in g(c)\setminus\{c\}$ and let $R_0$ contain only $(\code{deny},\{x\})$ at rank $d$, a pattern of literal characters that every rule language here can express. At steps $1,\dots,k-1$ the party is asked about calls $c_1,\dots,c_{k-1}$ whose saved patterns are disjoint from each other and from $g(c)$. None of them is denied or matched by an earlier saved rule, so each is prompted, and since no saved rule matches $c$, the party is asked about $c$ at step $k$. The $k$-th answer adds $(\code{allow},g(c))$ at rank $\pi_k>d$, and $x\in\M(R_0)$ is then decided by an allow rule.
\end{proof}

\paragraph{Wildcard Semantics} Claude Code matches \code{*} and \code{?} as Proposition~\ref{prop:conp} assumes: on build 2.1.314, a deny rule \code{Read(./s?.txt)} refused \code{s1.txt} and allowed
\code{s12.txt}, \code{s.txt}, and \code{t1.txt}, and \code{Read(./s*.txt)} refused all but \code{t1.txt}, in three runs each.

\paragraph{Exhaustive Check of Theorem~\ref{thm:saved}} The search of Section~\ref{sec:predict} covers a model with 3 calls, 3 ranks, every operator policy of at most 2 rules, every widening $g$, every sequence of ranks, and every run of up to 3 answers. It checks all 12,096 combinations and finds a lost restriction in exactly the 8,946 that the theorem calls unsafe.

\paragraph{Coverage Test} A runtime that wants the extra admissions of Theorem~\ref{thm:local}(ii) must decide whether the remaining deny rules cover a deleted one, which Proposition~\ref{prop:conp} makes coNP-hard for path rules. A filter can instead check that every deny rule before the update is covered by a single deny rule after it, in each policy that a session loads. This test runs in $O(|p|\,|q|)$ time, is exact for command rules, and can only make a filter refuse more for path rules. It answers yes only if it can split $p$ into consecutive pieces, one for each token of $q$, with each piece inside its token. A literal of $q$ takes the same literal, a \code{**} of $q$ takes any piece, and in path rules, which follow gitignore semantics, a \code{*} of $q$ takes literals other than \code{/} and single \code{*}s. Every string that $p$ matches then splits the same way, so $q$ matches it. A dynamic program finds such a split, and with one kind of wildcard it exists exactly when $q$ matches the string $p'$ of Proposition~\ref{prop:conp}. On every input of Section~\ref{sec:repaireval}, this test and one that reads every \code{*} as any string give the same results.

\section{Research Ethics}
\label{app:disclosure}

All experiments ran on our own machines against released software, local copies of it, or our own accounts with model providers, and we patched only local copies of Claude Code's binary to test the repair. We do not name the open-source relays that forward updates. The security policy of opencode describes its permission system as a usability feature rather than a security boundary, and its development branch has since adopted a safe way of saving answers. The deletion of Section~\ref{sec:runs} still succeeds on build 2.1.327, the latest release at the time of writing.

\end{document}